\documentclass{article}
\usepackage{graphicx} %
\usepackage[dvipsnames, table, hideerrors]{xcolor} %
\definecolor{LightBlue}{rgb}{0.5, 0.6, 0.9}
\definecolor{one}{RGB}{51,34,136}
\definecolor{two}{RGB}{17,119,51}
\definecolor{three}{RGB}{136,34,85}
\usepackage{fullpage}
\usepackage{libertine}
\usepackage[bb=boondox]{mathalfa}
\usepackage{authblk}

\usepackage{amssymb, amsmath, amsthm}

\usepackage{wrapfig}

\usepackage{lineno}
 \usepackage{algorithm}
 \usepackage{algorithmic}

\usepackage[pagebackref]{hyperref}
\hypersetup{
    colorlinks=true,
    linkcolor=red, %
    citecolor=blue, %
    filecolor=cyan, %
    urlcolor=one, %
}

\renewcommand*{\backref}[1]{}
\renewcommand*{\backrefalt}[4]{
    \ifcase #1 (Not cited.)
    \or        (#2)
    \else      (#2)
    \fi
}
\usepackage[capitalize]{cleveref}

\newcommand{\Lap}{\mathrm{Lap}}
\newcommand{\eps}{\ensuremath{\epsilon}}
\newcommand{\thresh}{\mathrm{Thresh}}
\newcommand{\densLap}[1]{f_{\mathrm{Lap}}(#1)}
\newcommand{\out}{\mathrm{out}}
\newcommand{\generalmodel}{general model}
\newcommand{\simplemodel}{``likes"-model}

\newcommand{\inp}{x}

\newcommand{\err}{\mathrm{err}}
\newcommand{\unisize}{d}

\newcommand{\countdist}{\textsc{CountDistinct}}
\newcommand{\outdet}{output-determined}
\newcommand{\Alg}{\mathcal{A}}
\newcommand{\indfun}{\mathbb{1}}
\newcommand{\Sk}{S_K}
\newcommand{\Skj}{S_{K_j}}

\newtheorem{theorem}{Theorem}
\newtheorem{lemma}{Lemma}
\newtheorem{corollary}{Corollary}

\newtheorem{fact}{Fact}
\newtheorem{definition}{Definition}

\makeatletter
\renewcommand\refname{References (with cited pages)}
\renewcommand\@biblabel[1]{\textcolor{two}{\textbf{\small #1}}}

\makeatother

\begin{document}

\title{Private Counting of Distinct Elements in the Turnstile Model \\ and Extensions}
\author[1]{Monika Henzinger}
\author[2]{A. R. Sricharan}
\author[3]{Teresa Anna Steiner}
\affil[1]{Institute of Science and Technology, Klosterneuburg, Austria}
\affil[2]{Faculty of Computer Science, Doctoral School Computer Science, University of Vienna, Austria}
\affil[3]{Technical University Denmark}
\date{}

\maketitle

\begin{abstract}
Privately counting distinct elements in a stream is a  fundamental data analysis problem with many applications in machine learning.
In the turnstile model, Jain~et~al.~[NeurIPS2023] initiated the study of this problem  parameterized by the maximum flippancy of any element, i.e., the number of times that the count of an element changes from 0 to above 0 or vice versa.  They give an item-level $(\epsilon,\delta)$-differentially private algorithm whose additive error is tight with respect to that parameterization. In this work, we show that a very simple algorithm based on the sparse vector technique achieves a tight additive error for item-level $(\epsilon,\delta)$-differential privacy and item-level $\epsilon$-differential privacy with regards to a different parameterization, namely the sum of all flippancies. Our second result is a bound which shows that for a large class of algorithms, including all existing differentially private algorithms for this problem, the lower bound from item-level differential privacy extends to event-level differential privacy. This partially answers an open question by Jain~et~al.~[NeurIPS2023]. \end{abstract}
\section{Introduction}
Counting distinct elements in a stream is a fundamental data analysis problem that is widely studied \cite{DBLP:journals/jcss/FlajoletM85,hyperloglog,DBLP:conf/pods/KaneNW10,DBLP:conf/kdd/KarppaP22,DBLP:conf/pods/WangP23} and has many applications \cite{akella2003detecting,DBLP:journals/ton/EstanVF06,DBLP:conf/edbt/MetwallyAA08,baker2019dashing,DBLP:journals/compnet/WeedageLS21,DBLP:conf/noms/ClemensSGH23}, including network analysis \cite{DBLP:journals/compnet/WeedageLS21} and detection of denial of service attacks \cite{akella2003detecting,DBLP:conf/noms/ClemensSGH23}. If the data includes sensitive information, the essential challenge is to give accurate answers while providing privacy guarantees to the data owners. Differential privacy is the de-facto standard in private data analysis and is widely employed both in research and in industry. In the insertions-only model, the problem of counting distinct elements while preserving differential privacy is well-studied \cite{DBLP:conf/icdt/BolotFMNT13,DBLP:conf/innovations/EpastoMMMVZ23,DBLP:conf/innovations/Ghazi0NM23}.

Recent work by Jain, Kalemaj, Raskhodnikova, Sivakumar, and Smith \cite{jain2023counting} (which was concurrent with an earlier version of the results presented in this paper, see \cite[Section 5]{henzinger2023differentially})
initiated the study of this problem in the more general turnstile model. They give an algorithm which is \emph{item-level}, $(\epsilon,\delta)$-differentially private and analyze the additive error parameterized in the {\em maximum flippancy} of any element, i.e., the number of times that the count of an element changes from 0 to above 0 or vice versa. They also give lower bounds which show that the additive error of the algorithm is tight for item-level differential privacy (up to log factors) with respect to their parameterization. There is still a gap for event-level differential privacy, which is posed as an open question. The algorithm is based on several instantiations of the binary tree mechanism.

In this paper we show that a simple algorithm based on the sparse vector technique achieves a tight additive error (up to log factors) for item-level $(\epsilon,\delta)$-differential privacy and item-level $\epsilon$-differential privacy, with regards to a different parameterization, namely the {\em total flippancy}, i.e., the sum of the flippancies of all elements. The additive error depends polynomially on the total flippancy with a smaller exponent than the exponent of the maximum flippancy in the additive error in \cite{jain2023counting}. Thus, if there are few elements in total, or few elements which change their count from 0 to above 0 or vice versa, then our algorithm achieves a better additive error. Additionally, we give  is a reduction which shows that for a large class of algorithms, including all existing differentially private algorithms for this problem, the lower bound from item-level differential privacy extends to event-level differential privacy. This is a step towards answering the open question posed in \cite{jain2023counting}. %

\subsection{Problem Definition}
More formally, we assume there are $d$ different items, and our goal is to maintain a multiset of them and to determine at each time step how many of them are currently at least once in the multiset, i.e., the number of distinct elements in the multiset. The update operations are modeled as follows: The input at every time step is a $\unisize$-dimensional vector $x^t \in \{-1, 0, 1\}^d$, such that $x^t_i=1$ if element $i$ gets inserted at time $t$, $x^t_i=-1$ if element $i$ gets deleted at time $t$, and $x^t_i=0$ otherwise. Note that this means that we allow \emph{multiple non-zero entries} in $x^t$, corresponding to \emph{multiple updates} at every time step. However, the lower bound also extends to the case where we assume that at most one element may be inserted or deleted at any time step, i.e., $||x^t||_1\leq 1$, which we call \emph{singleton update streams}. At every time step $t$, we want to output the number of distinct elements in the multiset. By our definition of the input stream, an element $i$ is present at time $t$ if and only if $\sum_{t'\leq t}x_i^{t'}>0$.
\begin{definition}[\countdist]\label{def:intro}
    Let $x^1,x^2,\dots, x^T$ be an input stream with $x^t\in \{-1,0,1\}^{d}$ for all $t\in[1,T]$. We define $\countdist(x)^t=\sum_{i=1}^d \indfun(\sum_{t'\leq t}\inp_i^{t'}>0)$, where $\indfun(E)$ is the indicator function that is $1$ if $E$ is true and $0$ otherwise. Then, the $\countdist$ problem is to output $\countdist(x)^t$ at all time steps $t$. The \emph{error} of \countdist\ is defined to be the maximum additive error over all time steps.
\end{definition}
In this paper, we consider two privacy notions: \emph{event-level} differential privacy, and \emph{item-level} differential privacy. They differ in their definition of neighboring input streams. Two input streams $x$ and $y$ are \emph{event-level neighboring}, if there exists a time step $t^{*}$ and an item $i^{*}\in  [1,d]$ such that we have $x_i^t=y_i^t$ for all $(i,t)\neq (i^{*},t^{*})$. That is, two event-level neighboring streams may differ in {\em  at most one item in at most one update operation}. Two input streams $x$ and $y$ are {\em item-level neighboring}, if there exists an item $i^{*}\in [1,d]$ such that $x_i^t=y_i^t$ for all $t$ and for all $i\in[1,d] \setminus \{i^*\}$. That is, two item-level neighboring streams may differ in \emph{all update operations related to one item}.

Finally, we consider two models regarding the input stream.
In the \emph{\generalmodel}  the counts for any item at any  time step $t$ is given by $\sum_{t'\leq  t} x_i^{t'}$, which can be any integer in $[-t,t]$ and
we only care about whether $\sum_{t'\leq  t} x_i^{t'}$ is larger than zero or not.
In the \emph{\simplemodel}\footnote{The name was chosen as it models the count of ``likes'' on a social media website, as motivated by \cite{jain2023counting}.}
for every item $i$ at any time step $t$, it must hold that $\sum_{t'\leq  t} x_i^{t'} \in \{0,1\}$, i.e., the multiset is a set. Said differently, an item can only be inserted if it is absent in the set and it can only be deleted when it is present.

\subsection{Summary of Results}
In this paper, we give new upper and lower bounds for \emph{item-level} differential privacy, parameterized in the \emph{total flippancy} $K$, which is defined as the total number of times any item switches from a non-zero count to a zero count, or vice versa. In detail, let $f^t(x_i)=\indfun(\sum_{t'\leq t} x_i^{t'}>0)$. The total flippancy is formally defined  as $K=\sum_{i=1}^d \sum_{t=2}^T\indfun(f^t(x_i)\neq f^{t-1}(x_i))$. Note that in the \simplemodel, the total flippancy is equal to the total number of updates.
As $\countdist(x)^t=\sum_{i=1}^d f^t(x_i)$, it follows that $K$ is an upper bound on the number of changes in $\countdist(x)$ over time. %

\paragraph*{Upper Bounds} As our first main result, we give algorithms solving \countdist\ while providing item-level differential privacy, which work in the \generalmodel\ (thus also in the \simplemodel). In the following, %
we state the exact bounds for \emph{given $K$}. If $K$ is not given to the algorithm, the error bounds worsen by at most a $\ln^2 K$ factor.

\begin{theorem}\label{thm:upperbound_intro}
   Let $d$ be a non-zero integer, $\beta>0$, $K$ be a known upper bound on the total flippancy, and let $T$ be a known upper bound on the number of time steps. Then there exists
  \begin{enumerate}
      \item

    an item-level $\epsilon$-differentially private algorithm for  the \countdist\ problem in the general model with additive error $O(\min(d,K,\sqrt{\epsilon^{-1}K\ln (T/\beta)},\epsilon^{-1}T\log(T/\beta))$ with probability at least $1-\beta$ at all time steps simultaneously, for any $\epsilon>0$ and $\beta \in (0,1)$; %

    \item an item-level $(\epsilon,\delta)$-differentially private algorithm for  \countdist\  in the general model with additive error $O(\min(d,K,\left(\epsilon^{-2}K\ln(1/\delta)\ln^2(T/\beta)\right)^{1/3},
\epsilon^{-1}\sqrt{T\ln(1/\delta)\log(T/\beta)})$  with probability at least $1-\beta$ at all time steps simultaneously, for any $\delta \in(0,1)$, $\epsilon \in (0,1)$, and $\beta \in (0,1).$ %
  \end{enumerate}
   \end{theorem}
As our lower bounds (discussed below) show, our bounds for $\epsilon$-differential privacy  are \emph{tight}, if  $K$ is known and $K\leq T$. If $K>T$, we incur at most an extra $\ln T$ factor, and if $K$ is not known, we incur extra $\ln K$ factors (see \cref{upper_bound_unknownK}). For $(\epsilon,\delta)$-differential privacy, the upper bounds are tight up to $\ln T$, $\ln K$ and $\ln(1/\delta)$ factors.

\paragraph*{Lower bounds} We complement our upper bounds by almost tight lower bounds on the additive error which hold for any item-level differentially private algorithm in the \simplemodel. As this is the ``more restricted'' of the two models, the lower bounds also carry over to the \generalmodel. For $\epsilon$-differential privacy, our lower bound follows from a packing argument. %

\begin{theorem}[Simplified version of Theorem~\ref{thm:lower_eps}]\label{thm:intro_lower_eps}
For any $L\leq T$, there exists an input stream $x$ of $d$-dimensional vectors from $\{-1,0,1\}^d$, which is valid in the \simplemodel, with length $T$ and flippancy $K=\Theta(L)$, such that any item-level, $\epsilon$-differentially private algorithm for \countdist\ must with constant probability have an error at least $\Omega(\min(d,K,\sqrt{\epsilon^{-1}K\max(\ln(T/K),1)}))$.
\end{theorem}
The lower bound above also holds for singleton updates. When multiple updates are allowed, then $K$ could potentially be larger than $T$. In that case, Theorem~\ref{thm:lower_eps} in \cref{sec:lowerbound} shows that for any $T\leq L\leq dT$, there exists a stream with flippancy $K=\Omega(T)$, $K=O(L)$, such that any item-level, $\epsilon$-differentially private algorithm to the \countdist\ problem must have error at least $\Omega(\min(d,\epsilon^{-1}T,\sqrt{\epsilon^{-1}L\max(\ln(T/L),1)}))$ with constant probability. Since $K = O(L)$, this gives a lower bound of $\Omega(\min(d,\epsilon^{-1}T,\sqrt{\epsilon^{-1}K\max(\ln(T/K),1)}))$. For $(\epsilon,\delta)$-differential privacy, we can use a similar strategy as in \cite{jain2023counting} to get the following bounds:
\begin{theorem}[Simplified version of Theorem~\ref{thm:lowerbound_epsdel}]\label{thm:lowerbound_epsdel_intro}
    Let $\epsilon,\delta\in(0,1]$.
    Let $K, T$ be sufficiently large parameters. There exists a dimension $d \in \mathbb N$ and an input stream $x$ of $d$-dimensional vectors from $\{-1,0,1\}^d$ of length $T$ with flippancy at most $K$ which is valid in the ``likes''-model, such that any item-level, $(\epsilon,\delta)$-differentially private algorithm for the \countdist\ problem must have error at least $\Omega\left(\eps^{-1} \cdot \min\left(\frac{\sqrt{T}}{\log T},\frac{(K\epsilon)^{1/3}}{\log (K\epsilon)}\right)\right)$ with constant probability.
\end{theorem}
Note that this lower bound holds for the case where \emph{multiple insertions} are allowed in every time step. In Theorem~\ref{thm:lowerbound_epsdel} we also give a lower bound of  $\Omega\left(\frac{K^{1/3}}{\epsilon\log K}\right)$ for singleton-updates.

\paragraph*{Time and Space Complexity}
Our main algorithm (achieving the $O(\sqrt{\epsilon^{-1} K \ln T})$ error bound for $\epsilon$-differential privacy and $O\left((K\ln(1/\delta)\ln^2 T)^{1/3}/{\epsilon^{2/3}}\right)$ error bound for $(\epsilon,\delta)$-differential privacy) can be implemented using constant time per update, assuming that drawing from a Laplace distribution takes constant time. Specifically, the total running time is $O(\# \textnormal{updates} + S_K t_{\Lap})$, where $t_{\Lap}$ is the time to draw one Laplace random variable, $S_K=O(\sqrt{K\epsilon/\ln T}+1)$ for $\epsilon$-dp, and $S_K=O\left(\left(\frac{K\epsilon}{\sqrt{\ln(1/\delta)}\ln(T/\beta)}\right)^{2/3}\right)$ for $(\eps, \delta)$-dp. The algorithm uses $O(d)$ words of space. The only information our algorithm needs to store are the true counts for each item, plus a constant number of words of extra information. This holds even for the case where $K$ is unknown, since we \emph{sequentially} run our known $K$ algorithm with increasing guesses for $K$.

\paragraph*{Comparison to the recent work by  Jain, Kalemaj, Raskhodnikova, Sivakumar, and Smith~\cite{jain2023counting}.}

\begin{table*}
\begin{centering}
\begin{tabular}{|c|c|c|c|c|}
    \hline & Item-level $\epsilon$-dp  & Item-level $(\epsilon,\delta)$-dp & Event-level $\epsilon$-dp &  Event-level $(\epsilon,\delta)$-dp \\\hline\hline
     \generalmodel & & $O(\min(\sqrt{w}, T^{1/3}))$ & & $O(\min(\sqrt{w}, T^{1/3}))$\\ ~\cite{jain2023counting} & $\Omega(\min(w, \sqrt{T}))$  & $\Omega(\min(\sqrt{w},T^{1/3})$& & {$\Omega(\min(\sqrt{w}, T^{1/4}))$}\\ \hline
     \simplemodel & & $O(\min(\sqrt{w}, T^{1/3}))$ & & $O(\min(\sqrt{w}, T^{1/3}))$\\  ~\cite{jain2023counting} & $\Omega(\min(w, \sqrt{T}))$ & $\Omega(\min(\sqrt{w},T^{1/3}))$ & & \\ \hline
    \generalmodel &\color{blue} $O(\sqrt{K})$ & \color{blue}$O(K^{1/3})$ & \color{blue}$O(\sqrt{K})$ & \color{blue}$O(K^{1/3})$\\ ~this work& \color{blue}$\Omega(\sqrt{K})$ &\color{blue} $\Omega(K^{1/3})$  & \color{blue}$\Omega(\min(w,\sqrt{K}))^{*}$ & \color{blue} $\Omega(\min(\sqrt{w},K^{1/3}))^{*}$ \\ \hline
     \simplemodel  & \color{blue}$O(\sqrt{K})$ & \color{blue}$O(K^{1/3})$ & \color{blue}$O(\sqrt{K})$ & \color{blue}$O(K^{1/3})$ \\ ~this work &\color{blue}$\Omega(\sqrt{K})$ & \color{blue}$\Omega(K^{1/3})$ & & \\ \hline
    \simplemodel & & & $O(1)$ & $O(1)$ \\ \hline
\end{tabular}
\caption{Comparison of our results  (in blue) and the results in \cite{jain2023counting} for the different models. $K$ denotes the total flippancy and $w$ denotes the maximum flippancy of an input stream $x$. For simplicity of exposition, we consider singleton insertions and omit factors polynomial in $\ln T$, $\ln(1/\delta)$, and $\epsilon^{-1}$. The bounds marked with $^*$ hold for \emph{output dependent} algorithms (see the discussion before Theorem~\ref{thm:intro_event-level} for details). The bounds in the last line follow from a simple application of a continual counting algorithm on the difference sequence.
}
\end{centering}
\end{table*}

In recent work, \cite{jain2023counting} considered the \countdist\ problem with a similar, but with a different parameterization.
In \cite{jain2023counting}, they parameterize the additive error in the \emph{maximum flippancy}, i.e., they parameterize on $w_x=\max_{i\in[d]}(\sum_{t=2}^T \indfun(f^t(x_i)\neq f^{t-1}(x_i))$. Recall that $K$ denotes the total flippancy of a stream $x$ and note that $w_x\leq K\leq d\cdot w_x$. \cite{jain2023counting} consider only streams with singleton updates and give algorithms for item-level, $(\epsilon,\delta)$-differential privacy in the \generalmodel, with an error bound of $\tilde{O}\left(\min\left((\sqrt{w_x}\log T + \log^3 T)\cdot \frac{\sqrt{\log(1/\delta)}}{\epsilon},\right.\right.$ $\left. \left. \frac{(T\log(1/\delta))^{1/3}}{\epsilon^{2/3}},T\right)\right)$\footnote{For simplicity of exposition, we use $\tilde{O}(X)$ to denote $O(X\cdot \mathrm{polylog}(X)$)}.
In comparison, our bounds in this setting are $\tilde{O}\left(\min\left(\frac{(K\ln(1/\delta)\ln^2 T)^{1/3}}{\epsilon^{2/3}}\right),K\right)$. Note that $K\leq T$ for singleton updates, and thus, our upper bounds recover their second and third bound up to a $\ln^{2/3}T$ factor. Furthermore, ignoring polynomial factors in $\log T$, $\log(1/\delta)$ and $\epsilon^{-1}$, their bound is $O(\sqrt{w_x})$ while ours is $O(K^{1/3})$. Thus, if (roughly) $K <  w_x^{3/2}$, our algorithm outperforms theirs. Specifically, if $d\leq \sqrt{w_x}$ or if there are only few items with high flippancy, we expect our algorithm to do better. In cases where the flippancy is well-distributed, i.e., many items have a similar flippancy, and $d\geq \sqrt{w_x}$, we expect the algorithm in \cite{jain2023counting} to perform better.

In terms of space and time complexity, their algorithm, like ours, needs to maintain a count for each element. Thus, the space in terms of words is $\Omega(d)$. On top of that, they run a variant of the binary tree mechanism, which depending on the implementation, uses $\Omega(\log T)$ space. In their final solution, they actually run $\log T$ copies of the binary tree mechanism in parallel, bringing their space consumption to $O(d+\log^2 T)$ words. Thus, the space of our algorithm is an additive $\log ^2 T$ term better, which can be crucial for large streams. In terms of time complexity, each of the binary tree mechanism needs to draw $\Omega(T\log T)$ independent Laplace noises, thus their time complexity is at least $\Omega(T\log^2 T t_{\Lap})$, where $t_{\Lap}$ is the time it takes to draw a Laplace noise. Also here, our algorithm is more efficient.

In terms of lower bounds, for item-level, $\epsilon$-differential privacy in the \simplemodel,  \cite{jain2023counting} give a lower bound of $\Omega(\min(\epsilon^{-1}w,\sqrt{\epsilon^{-1}T}, T))$ for streams of maximum flippancy at most $w$. For $(\epsilon,\delta)$-differential privacy, they give a lower bound of $\tilde{\Omega}(\min(\epsilon^{-1}\sqrt{w},\epsilon^{-2/3}T^{1/3},T))$ for item-level privacy in the \simplemodel, and a lower bound of $\tilde{\Omega}(\min(\epsilon^{-1}\sqrt{w},\epsilon^{-3/4}T^{1/4},T))$ for event-level privacy in the \generalmodel, for streams of maximum flippancy at most $w$. Their upper bounds in the item-level setting match their lower bounds up to factors polynomial in $\log T$ and $\log(1/\delta)$. For event-level in the \generalmodel, there is a gap for $\sqrt{T}\leq w\leq T^{2/3}$, and closing this gap was posed as an explicit open question in \cite{jain2023counting}.\footnote{Note that this gap only exists if at most one update per time step is allowed - if many (e.g. up to $d$ many) updates are allowed in each time step, then the lower bound proof for event-level privacy in the general model from \cite{jain2023counting} can be used to show a lower bound of $\Omega(\sqrt{T})$, for constant $\epsilon$ and $\delta$.} As our second main result, we make a step towards closing this gap, which we explain below.

\paragraph*{Reduction from item-level, \simplemodel\ to output-dependent event-level, \generalmodel}
All the upper bounds mentioned so far hold for \emph{item-level} differential privacy. As our upper bounds hold in the \generalmodel\ and our lower bounds hold in the \simplemodel, we can conclude that for item-level privacy, the \simplemodel\ and the \generalmodel\ are roughly equally hard. \cite{jain2023counting} arrived at this conclusion as well, albeit with a different parameterization.

However, for {\em event-level} differential privacy, the picture is different: for the ``likes''-model, a very simple algorithm gives an error of $O(\epsilon^{-1}\mathrm{polylog}(T))$ with constant probability. To see this, define the \emph{difference sequence} for the \countdist\ problem as $\mathrm{diff}^t(x)= \countdist(x)^t-\countdist(x)^{t-1}$ for $t>1$. As can be easily seen, $(\mathrm{diff}^t(x))_{t>1}$ and $(\mathrm{diff}^t(y))_{t>1}$ differ by at most 1 in at most one time step $t$ for any event-level neighboring streams $x$ and $y$ in the \simplemodel. Thus, applying a standard continual counting algorithm gives the claimed error, as shown for ``well-behaved'' difference sequences in general in \cite{DBLP:conf/esa/FichtenbergerHO21}.

For event-level differential privacy and the \emph{\generalmodel}\ however, the best known algorithms are the algorithms for item-level differential privacy in this paper and \cite{jain2023counting}. \cite{jain2023counting} also present lower bounds for event-level differential privacy in the \generalmodel\  which, however, leave a gap for certain parameter settings. Closing that gap was explicitly posed as an open question in~\cite{jain2023counting}. We make a step towards closing that gap, by noting that all existing differentially private algorithms for the \countdist\ problem in \emph{any} model share the following property: If $\countdist(x)=\countdist(y)$ for any two input streams $x$ and $y$, then the output distributions of the algorithms are equal. That is, any two streams which produce the same true output, will have the same output distributions. We call such algorithms \emph{output-determined}. We show that if we only consider output-determined algorithms for \countdist, then achieving \emph{event-level differential privacy in the \generalmodel\ is just as hard as item-level differential privacy for the \simplemodel}. Thus our above lower bounds also apply to such algorithms. In particular, this shows that if one were trying to close the gap for event-level differential privacy in the \generalmodel, one needs to find an algorithm which does not only depend on the true answers to \countdist.

\begin{theorem}[Simplified version of Theorem~\ref{thm:event-level}]\label{thm:intro_event-level}
Let $\epsilon > 0$ and $\delta\ge 0$.
    Let $\Alg_1$ be an event-level, $(\epsilon,\delta)$-differentially private, \outdet\ algorithm for \countdist\ that works in the \generalmodel\ and has error at most $\alpha$ for streams of length $T+1$ with probability $1-\beta$. Then there exists an item-level, $(2\epsilon,(1+e^{\epsilon})\delta)$-differentially private algorithm $\Alg_2$ for \countdist\ that works in the \simplemodel, and has error at most $\alpha$ for streams of length $T$ with probability $1-\beta$.
\end{theorem}

\paragraph*{Generalizations \& Applications}
While our algorithms are (nearly) tight for the \countdist\ problem, they are not tailored specifically to the problem and work in a more general setting as well. In particular, recall that $\countdist(x)^t=\sum_{i=1}^d f^t(x_i)$, where $f^t(x_i)=\indfun(\sum_{t'\leq t} x_i^{t'}>0)$. Now consider any real-valued function $Q$ on input streams $x_1,x_2,\dots,$ with $x_i \in  \{-1, 0, 1\}$. We use $Q^t(x)$ to denote $Q(x_1,\dots, x_t)$. Our algorithm works for any such function $Q$ such that the following two conditions are fulfilled:
      (1) \label{prop1} for any $x$ and $y$ which are neighboring, we have $|Q^t(x)-Q^t(y)|\leq 1$ for all time steps $t$, and %
    (2) \label{prop2} $\sum_{t=1}^T|Q^t(x)-Q^{t-1}(x)|\leq K$.
\begin{theorem}\label{thm:generalization}
    Let $Q$ be a function satisfying properties (1) and (2). Then there exists
    \begin{enumerate}
       \item

        an item-level $\epsilon$-differentially private algorithm for computing $Q$ with additive error $$O(\min(K, \sqrt{\epsilon^{-1}K\ln(T/\beta)})),\epsilon^{-1}T\log(T/\beta))$$ at all time steps with probability at least $1-\beta$, for any $\epsilon>0$;

        \item an item-level $(\epsilon,\delta)$-differentially private algorithm for computing $Q$ with additive error $$O(\min(K,(\epsilon^{-2}K\ln(1/\delta)\ln^2(T/\beta))^{1/3},\epsilon^{-1}\sqrt{T\ln(1/\delta)\log(T/\beta)}))$$ at all time steps with probability at least $1-\beta$, for any $\epsilon>0$, $\delta \in (0,1)$;
    \end{enumerate}
\end{theorem}
The extension to unknown $K$ also holds, with extra $\ln K$ factors as earlier.
Thus, for a continuous function $Q$ which has \emph{maximum} sensitivity 1 \emph{over all time steps}, we get a bound parameterized in the sum of all differences, i.e., the $L_1$-norm of the difference sequence.
While our results hold in the
{\em turnstile model} and the additive error is parametrized by the total flippancy, \cite{DBLP:conf/esa/FichtenbergerHO21} gave an $\epsilon$-differentially private mechanism with additive error $O(\Gamma \log^{3/2} \log(T/\beta))$ in the {\em insertions-only} or {\em deletions-only}
setting, where $\Gamma$ is the {\em continuous global sensitivity} which is the $L_1$-norm of the difference sequence of two neighboring inputs.

We can apply our algorithm to the problem considered in Fichtenberger et al.~\cite{DBLP:conf/esa/FichtenbergerHO21} of continuously counting high degree nodes under differential privacy, which counts the number of nodes with degree at least $\tau$, where $\tau$ is given and public. For user-level, edge-differential privacy (i.e., neighboring streams may differ in all updates of the same edge), they give a lower bound of $\Omega(n)$. %
Our algorithm gives new parameterized bounds for this problem: In particular, choosing $Q^t(x)=\frac{\textnormal{\# of high degree nodes}}{2}$, Theorem~\ref{thm:generalization} gives an error bound of roughly %
$O(\sqrt{K})$, under $\epsilon$-differential privacy, and roughly $O(K^{1/3})$ under $(\epsilon,\delta)$-differential privacy, where we ignore an additive factor of $O(\epsilon^{-1}\ln T\ln(1/\delta)\ln K)$. Note that $K$ can be as large as $T$, but for many applications, it could be much smaller: for example, in social networks, it has been shown that the degree distribution follows a power-law distribution, which implies that the set of high-degree nodes only changes infrequently. $K$ does not to be given to the algorithm.

\subsection{Algorithm Overview}

The main idea of our algorithm is to use the sparse vector technique first introduced by Dwork, Naor, Reingold, Rothblum, and Vadhan~\cite{DBLP:conf/stoc/DworkNRRV09} (the form we use it in can be found in Dwork and Roth~\cite{journals/fttcs/DworkR14}) on carefully chosen queries and with carefully chosen thresholds. The sparse vector technique can be described as follows: It is given a data set $x$, a sequence of $q$ queries, a threshold {\em Thresh}, and a stopping parameter $S$. It will process these queries sequentially, and for each of them answer ``yes'' or ``no'' depending on whether or not $q(x)$ is approximately (up to an additive error $\alpha$) above the threshold. It stops after it has answered ``yes'' $S$ times. Dwork and Roth~\cite{journals/fttcs/DworkR14} show that it is possible to design an $\epsilon$-differentially private algorithm achieving the above with $\alpha=O(\epsilon^{-1}S\log (q/\beta))$ with probability $1-\beta$, and an $(\epsilon,\delta)$-differentially private algorithm with $\alpha=O(\epsilon^{-1}\sqrt{S\log(1/\delta)}\log (q/\beta))$ with probability $1-\beta$. In the following discussion, we ignore $\epsilon^{-1}, \log(1/\delta), \log q$ and $\log(1/\beta)$ factors.

Our main idea is to note that the total flippancy $K$ can be seen as an upper bound on \emph{the total change in the output}, i.e., the sum of the absolute differences in the output in every time step. Our strategy is as follows: We start by estimating the number of distinct elements at the beginning of the stream. Then, we keep reporting this estimate until a significant change occurs in the true number of distinct elements. %
We track whether such a change has occured using the sparse vector technique. Once there has been a significant change, i.e., once the sparse vector technique answers ``yes'', we update the output. The goal now is to balance the additive error of the sparse vector technique with the error accumulated between updates. The error between updates is roughly {\em Thresh}; the error of the sparse vector technique is $\alpha$; and the total change of the output is bounded by $K$. To balance the two we set {\em Thresh} $=\Theta(\alpha)$. Furthermore we have to choose $S$ in a way that makes sure that the sparse vector technique does not abort before we have seen the entire stream. We can show that every time our sparse vector technique answers ``yes'', the change in output has been roughly {\em Thresh}. Thus it is enough to set $S>K/${\em Thresh}. As mentioned above, for $\epsilon$-differential privacy $\alpha$ (and, thus, {\em Thresh}) must depend linearly on $S$, which implies that  $S$ must be chosen to be $\Theta(\sqrt{K})$, giving an additive error of $O(\sqrt{K})$. For $(\epsilon,\delta)$-differential privacy,
 we have {\em Thresh}$ =\Theta(\alpha)=O(\sqrt{S})$.
This implies that $S^{3/2}$ must be $\Theta(K)$, i.e., $S=\Theta(K^{2/3})$. Thus the additive error is $O(K^{1/3})$.

Note that this requires that $K$ is known at the beginning of the algorithm. If $K$ is unknown, we run the above algorithm %
for exponentially increasing guesses of $K$ ($K=2,4,8,$ etc.). In particular, we run the algorithm for a guess of $K$, and if it terminates preemptively, we double our guess and repeat. Since we do not know beforehand how many instances are needed, in order to make sure the resulting algorithm is still $\epsilon$-differentially private, we run the $j$th instance with privacy parameter $\epsilon_j=O(\epsilon/j^2)$, such that $\sum_{j=1}^{\infty}\epsilon_j\leq \epsilon$. At the end of the algorithm, $j=\Theta(\ln K)$, therefore we incur an extra $\ln^2 K$ factor in the additive error. %

\section{Preliminaries}
We denote $\{1,\dots,n\}$ by $[n]$ and the input stream length by $T$, which is the number of time steps.

\paragraph*{Continual observation algorithm.}
An algorithm $A$ in the continual observation model gets an update at every time step $t \le T$, and produces an output $a^t=A(x^1,\dots,x^t)$ which is a function of $x^1$ to $x^t$; $A^T(x)=(a^1,a^2,\dots,a^T)$ denotes the sequence of outputs at all time steps.

\begin{definition}[Differential privacy \cite{Dwork2006}]\label{def:dp} A randomized algorithm $A$ is \emph{$(\epsilon,\delta)$-differentially private ($(\epsilon,\delta)$-dp)} if for all $S\in \mathrm{range}(A^T)$ and all $x,y$ neighboring
    \begin{align*}
        \Pr[A^T(x)\in S]\leq e^{\epsilon}\Pr[A^T(y)\in S]+\delta.
    \end{align*}
    If $\delta=0$ then $A$ is \emph{$\epsilon$-differentially private ($\epsilon$-dp)}.
    \end{definition}

\begin{definition}[Laplace Distribution]
The \emph{Laplace distribution} centered at $0$ with scale $b$ is the distribution with probability density function
$
\densLap{b}(x)=(2b)^{-1} \cdot \exp\left(-|x|/b\right).
$
We use $X\sim \Lap(b)$ or just $\Lap(b)$ to denote a random variable $X$ distributed according to $\densLap{b}(x)$.
\end{definition}

In our definitions below, we use $\chi$ to represent a generic universe of elements.

\begin{definition}[Sensitivity]
Let $f:\chi\rightarrow \mathbb{R}^k$. The $L_p$-sensitivity $\Delta_p$ is defined as
\[\max_{x\in \chi,y\in \chi, x\sim y}||f(x)-f(y)||_p,\]
where $x\sim y$ denotes that $x$ and $y$ are neighbouring.
\end{definition}

\begin{fact}[Theorem 3.6 in \cite{journals/fttcs/DworkR14}: Laplace Mechanism] \label{fact:Laplacemech} Let $f$ be any function $f:\chi\rightarrow \mathbb{R}^k$ with $L_1$-sensitivity $\Delta_1$. Let $Y_i\sim \Lap(\Delta_1/\epsilon)$ for $i\in[k]$. The mechanism defined as
$
A(x)=f(x)+(Y_1,\dots,Y_k)
$
satisfies $\epsilon$-differential privacy.
\end{fact}

\begin{fact}[Laplace Tailbound]\label{fact:laplace_tailbound}
If $X\sim \Lap(b)$, then
$
\Pr[|X|\geq t\cdot b]\leq e^{-t}.
$
\end{fact}

The following fact follows from Theorem A.1 in \cite{journals/fttcs/DworkR14}:
\begin{fact}[Gaussian Mechanism]\label{fact:gaussianmech} Let $f$ be any function $f:\chi\rightarrow \mathbb{R}^k$ with $L_2$-sensitivity $\Delta_2$. Let $Y_i\sim \mathcal{N}(0,\sigma^2)$ for $i\in[k]$, where $\sigma\geq \sqrt{2\ln(2/\delta)}\Delta_2/\epsilon$. The mechanism defines as
$
A(x)=f(x)+(Y_1,\dots,Y_k)
$
satisfies $(\epsilon,\delta)$-differentially privacy.
\end{fact}

\begin{fact}[Gaussian tailbound]\label{fact:gaussiantail}
If $X\sim\mathcal{N}(0,\sigma^2)$, then
$
\Pr[|X|\geq\sigma\sqrt{\ln(2/\beta)}]\leq \beta
$
\end{fact}

The following facts are respectively given by Theorem~3.16, 3.20 and Corollary 3.21 in \cite{journals/fttcs/DworkR14}.

\begin{fact}[Composition Theorem]\label{fact:composition_theorem} Let $A_1$ be an $(\epsilon_1,\delta_1)$-differentially private algorithm $A_1:\chi\rightarrow \mathrm{range}(A_1)$ and $A_2$ an $(\epsilon_2,\delta_2)$-differentially private algorithm $A_2:\chi\times\mathrm{range(A_1)}\rightarrow \mathrm{range}(A_2)$. Then $B:\chi\rightarrow \mathrm{range(A_1)}\times\mathrm{range}(A_2)$ defined as $B(x)=(A_1(x), A_2(x,A_1(x))$ is $(\epsilon_1+\epsilon_2,\delta_1+\delta_2)$-differentially private. \end{fact}

\begin{fact}[Advanced Composition]\label{fact:advanced_composition} Let $\epsilon,\delta,\delta'\geq 0$. Let $A_1$ be an $(\epsilon,\delta)$-differentially private algorithm $A_1:\chi\rightarrow \mathrm{range}(A_1)$ and $A_i$ be $(\epsilon,\delta)$-differentially private algorithms $A_i:\chi\times \mathrm{range}(A_{i-1})\rightarrow \mathrm{range}(A_i)$, for $2\leq i\leq k$.  Then the composition $B:\chi\rightarrow \mathrm{range}(A_1)\times\dots \times \mathrm{range}(A_k)$ defined as
\[
B(x)=(A_1(x), A_2(x,A_1(x)),\dots, A_k(x,A_{k-1}(x)))
\]
is $(\epsilon',k\delta+\delta')$-differentially private, where
$
\epsilon'=\sqrt{2k\ln(1/\delta')}\epsilon+k\epsilon(e^{\epsilon}-1).
$
\end{fact}

\begin{corollary}
Let $\epsilon^*,\delta,\delta'\geq 0$ and $\delta',\epsilon^*<1$. Let $A_1,\dots,A_k$ be as in Fact~\ref{fact:advanced_composition} with
\begin{align*}
\epsilon=\epsilon^*/(2\sqrt{2k\ln(1/\delta')}).
\end{align*}
Then the composition $B$ (defined as in Fact~\ref{fact:advanced_composition}) is $(\epsilon^*,k\delta+\delta')$-differentially private.
\end{corollary}

\section{Item-Level Algorithms in  General Model}\label{sec:upperbound}
In this section, we give algorithms which work for any input sequence in the \generalmodel, and thus also for input sequences that fulfill the conditions of the \simplemodel. The upper bounds on the additive error for $\epsilon$-differential privacy match the lower bounds in \cref{sec:lowerbound}, except for the $\log (T/\beta)$ factor in the case where $K> T$.
\subsection{Known Total Flippancy}\label{sec:totalknownflippancy}

We prove Theorem~\ref{thm:upperbound} in this section.
We give some intuition first on Algorithm~\ref{alg:monitoring}. The algorithm works by iteratively checking if the true number of distinct elements currently present (called $Q$) is ``far'' from the current output of our algorithm (called $\out$) using a sparse vector technique (SVT) instantiation. We start the algorithm by estimating $\out$ at the beginning of the stream (line \ref{line:monitoringFirstout}). Then, we keep outputting $\out$, while we track the difference between $\out$ and the true number of distinct elements $Q$ (line \ref{line:monitoringIf}). Once there has been a significant change, we update the output (line~\ref{line:monitoringUpdateout}).

There are two parameters of interest here. One is the number of times we update the output: we abort after $S_K$ updates happen (line~\ref{line:abort}). The other is the parameter $\thresh$, which determines how large the current error needs to be such that we satisfy the condition in line~\ref{line:monitoringIf}. The parameter $S_K$ goes into the error from composition, while the parameter $\thresh$ directly goes into the additive error bound.

The goal is to balance the error accumulated between updates (which is roughly $\thresh$), and the error from updating $\out$ privately (which is roughly $S_K$ for $\epsilon$-differential privacy, and roughly $\sqrt{S_K}$ for $(\epsilon,\delta)$-differential privacy due to composition). Additionally, we want to make sure our algorithm does not abort before having processed the entire stream. We show that every time SVT returns ``yes", the total flippancy in the stream has increased by at least $\Omega(\thresh)$. Since we know the total flippancy is bounded by $K$, in order to make sure that we do not abort preemptively, we choose $S_K$ such that $S_K\cdot \thresh \approx K$. Balancing the two error terms  yields an additive error of approximately $\sqrt{K}$ for $\epsilon$-differential privacy, and $K^{1/3}$ for $(\epsilon,\delta)$-differential privacy.

\begin{theorem}\label{thm:upperbound}
   Let $d$ and $T$ be non-zero integers, let $\beta>0$, and let $K$ be an upper bound on the total flippancy which is given. Let $T$ be a known upper bound on the number of time steps. Then there exists
  \begin{enumerate}
      \item

        an item-level $\epsilon$-differentially private algorithm for  the \countdist\ problem in the general model with error at most $O(\min(d,K,\sqrt{\epsilon^{-1}K\ln (T/\beta)},\epsilon^{-1}T\log(T/\beta))$ at all time steps with probability at least $1-\beta$ for $\epsilon>0$; %

      \item
        an item-level $(\epsilon,\delta)$-differentially private algorithm for  the \countdist\ problem in the general model with error $O\left(\min(d,K,\left(\epsilon^{-2}K\ln(1/\delta)\ln^2(T/\beta)\right)^{1/3},\right.$ $\left. \epsilon^{-1}\sqrt{T\ln(1/\delta)\log(T/\beta)}\right)$ at all time steps with probability at least $1-\beta$, for any $0<\delta<1$ and $0<\epsilon<1$. %
  \end{enumerate}
\end{theorem}

\begin{algorithm}[!ht]
\begin{algorithmic}[1]
\STATE {\bf Input:} Data Stream $x=x^1, x^2,\dots$, initial counts $c_1,\dots,c_d$ (default 0), parameters $\epsilon$, $\delta$ and $\beta$, stream length bound $T$, stopping parameter $\Sk\geq 1$
\STATE \textbf{if $\delta=0$ then} $\epsilon_1=\epsilon/(2\Sk)$
\STATE \textbf{if $\delta>0$ then } $\epsilon_1=\epsilon/(4\sqrt{2\Sk\ln(1/\delta)})$
\STATE $\mathrm{count}=1$
\STATE $\tau_1=\Lap(2/\epsilon_1)$
\STATE $\nu_1=\Lap(1/\epsilon_1)$

\STATE $Q=0$\label{line:setQ0}\;

\STATE $\out= Q+\nu_1$\label{line:monitoringFirstout}\;

\STATE $\mathrm{Thresh}=16\epsilon_1^{-1}(\ln(2T/\beta))$
\FOR{$t=1,\dots,$}
    \STATE $c_i=c_i+x_i^t$ for all $i\in[d]$

    \STATE  $Q= | \{ i \in [d] \mid c_i > 0 \} |$\label{line:setQt}\;

   \STATE $\mu_t=\Lap(4/\epsilon_1)$
    \IF{$|\out-Q|+\mu_t>\mathrm{Thresh}+\tau_{\mathrm{count}}$\label{line:monitoringIf}}

       \STATE $\mathrm{count}=\mathrm{count}+1$
        \STATE $\tau_{\mathrm{count}}=\Lap(2/\epsilon_1)$
        \STATE $\nu_{\mathrm{count}}=\Lap(1/\epsilon_1)$
       \STATE $\out=Q+\nu_{\mathrm{count}}$\label{line:monitoringUpdateout}\ENDIF

        \STATE \textbf{output} $\out$\;

        \STATE \textbf{if $\mathrm{count}\geq \Sk$\label{line:abort} then {Abort}}
\ENDFOR
\end{algorithmic}
\caption{\countdist, known $K$}
\label{alg:monitoring}
\end{algorithm}

\begin{proof}

The $O(\min(d,K))$ bound follows from the fact that the algorithm that outputs $0$ at every time step is $\epsilon$-differentially private and has error at most $\min(d,K)$ for any $\epsilon$. The third error bounds in the minimum for Theorem~\ref{thm:upperbound} are achieved by Algorithm~\ref{alg:monitoring}, as shown below. Since we assume here all parameters are known, one can compute the minimum of the three bounds and choose the algorithm accordingly.
The fourth bound in Theorem~\ref{thm:upperbound} follow by a direct application of the Laplace mechanism Fact~\ref{fact:Laplacemech} with $\Delta_1=T$ resp. Gaussian mechanism Fact~\ref{fact:gaussianmech} with  $\Delta_2 = \sqrt{T}$.

 The algorithm for our third bound, given in Algorithm \ref{alg:monitoring}, is based on the sparse vector technique, where $\Sk$ is a parameter dependent on $K$ that we choose suitably below. %
 We omit the proof of the following lemma, since it follows from well-known techniques (Sparse Vector Technique~\cite{DBLP:conf/stoc/DworkNRRV09, journals/fttcs/DworkR14}, Laplace mechanism (Fact~\ref{fact:Laplacemech}) and composition theorems (Facts~\ref{fact:composition_theorem} and~\ref{fact:advanced_composition})).
     \begin{lemma}\label{lem:monitoring_private}
        For $\delta=0$ and any $\epsilon>0$, Algorithm \ref{alg:monitoring} is $\epsilon$-differentially private. For $0<\epsilon<1$ and $0<\delta<1$, Algorithm \ref{alg:monitoring} is $(\epsilon,\delta)$-differentially private.
    \end{lemma}

We show the claimed accuracy bound using the following lemma.

    \begin{lemma}\label{lem:monitoring_accurate}
    For $\delta=0$, for any time step $t$ before the algorithm aborts, we have that the maximum error up to time $t$ is at most $O(\epsilon^{-1}S_K\ln(T/\beta))$. Setting $\Sk=\sqrt{K\epsilon/(18\ln(T/\beta))}+1$, with probability at least $1-\beta$, Algorithm \ref{alg:monitoring} does not abort before having seen the entire stream, and has error at most $O(\sqrt{\epsilon^{-1}K\ln (T/\beta)}+\epsilon^{-1}\ln(T/\beta))$.
    For $\delta>0$, for any time step $t$ before the algorithm aborts, we have that the maximum error up to time $t$ is  $O(\epsilon^{-1}\sqrt{S_K\ln(1/\delta)}\ln(T/\beta))$.
   Setting $\Sk=\left(\frac{K\epsilon}{36 \sqrt{\ln(1/\delta)}\ln(T/\beta)}\right)^{2/3}+1$,  with probability at least $1-\beta$, Algorithm \ref{alg:monitoring} does not abort before having seen the entire stream, and has error at most $O\left(\left(\epsilon^{-2}K\ln(1/\delta)\ln^2(T/\beta)\right)^{1/3}\right.$ $\left.+\epsilon^{-1}\sqrt{\ln(1/\delta)}\ln(T/\beta)\right)$.
    \end{lemma}%
    \begin{proof}

  Note that at every time step $t$ in Algorithm~\ref{alg:monitoring}, we set $Q=\sum_{i=1}^d f^t(\inp_i)$. Let $\alpha=(8/\epsilon_1)\ln(2T/\beta)= (1/2) \cdot\mathrm{Thresh}$.
 By Laplace tailbounds (Fact \ref{fact:laplace_tailbound}), at every time step $t$:

     (a) $|\tau_{{\ell}}|\leq (2/\epsilon_1)\ln(2T/\beta) = \alpha/4$ with probability at least $1-\beta/(2T)$, where $\ell$ is the value of variable $\mathrm{count}$ at time step $t$, and

     (b) $|\mu_t|\leq (4/\epsilon_1)\ln(2T/\beta)= \alpha/2$ with probability at least $1-\beta/(2T)$.

\noindent
Thus, with probability $ \ge 1-\beta$, we have at all time steps $t$ simultaneously:%

     (i) Whenever the condition in line~\ref{line:monitoringIf} is true at time $t$, then $|\out-\sum_{i\in[d]}f^t(\inp_i)|> \mathrm{Thresh}-3\alpha/4=5\alpha/4$, and

     (ii) Whenever the condition in line~\ref{line:monitoringIf} is false at time $t$, then $|\out-\sum_{i\in[d]}f^t(\inp_i)| \le\mathrm{Thresh}+3\alpha/4 <3\alpha$.

Further, the random variable $\nu_{\ell}$ for $\ell\in[\Sk]$ is distributed as Lap$(1/\epsilon_1)$ and is added to $\sum_{i\in[d]}f^t(\inp_i)$ at every time step $t$ where $\out$ is updated. By the Laplace tail bound (Fact \ref{fact:laplace_tailbound}), $\nu_{\ell}$
is bounded for \emph{all} $\ell\in[\Sk]$ by $\epsilon_1^{-1}\ln(\Sk/\beta)\leq\alpha/8 $ with probability at least $1-\beta$.
Altogether, all of these bounds hold simultaneously with probability at least $1-2\beta$. We condition on all these bounds being true.

Assume the algorithm has not terminated yet at time $t$ and let $\out$ be the value of variable $\out$ at the beginning of time $t$.
Let $p_{\ell}$ be the last time step at which the value of $\out$ was updated. It holds that $|\out- \sum_{i\in[d]}f^{p_{\ell}}_i(x)| = |\nu_{\ell}|\leq \alpha/8$.
If the condition in line~\ref{line:monitoringIf} is true at time $t$, then
\begin{equation*}
\left|\sum_{i\in[d]}f^{p_{\ell}}_i(x)-\sum_{i\in[d]}f^t(\inp_i)\right|
\geq \left|\sum_{i\in[d]}f^t(\inp_i)-\out\right| -
\left|\out- \sum_{i\in[d]} f^{p_{\ell}}_i(x) \right|
\geq 5\alpha/4-\alpha/8
= 9\alpha/8.
\end{equation*}
Thus, between two time steps where the value of $\out$ is updated, there is a change of at least $9\alpha/8$ in the sum value, i.e.,~the value of $f^{t}(x_i)$ has changed at least once for  $\ge 9\alpha/8$ different items $i$.
Since $K=\sum_{i=1}^{\unisize}\sum_{t=2}^T \indfun(f^t(x_i)\neq f^{t-1}(x_i))$, to guarantee (under the noise conditions), that the algorithm does not terminate before we have seen the entire stream, it suffices to choose $\Sk$ where $\Sk>K/(9\alpha/8)$.

For $\delta=0$, we have $\alpha=(8/\epsilon_1)\ln(2T/\beta)=(16\Sk/\epsilon)\ln(2T/\beta)$, thus we have to choose $\Sk$ to be at least $K\epsilon/(18\Sk\ln(2T/\beta))$. Choosing
$\Sk = \lfloor \sqrt{K\epsilon/(18\ln (2T/\beta))}\rfloor +1$
fulfills this condition. A similar calculation show that for $\delta>0$, choosing
$\Sk=\left(\frac{K\epsilon}{36\sqrt{\ln(1/\delta)}\ln(T/\beta)}\right)^{2/3}+1$
fulfills this condition.

Now %
consider any time step $t$ and let $\out$ be the output at time $t$, i.e., the value \emph{after} processing time step $t$. If the condition in line~\ref{line:monitoringIf} is false, we showed above that $|\out- \sum_{i\in[d]}f^t(\inp_i)|<3\alpha$. If the condition is true at time $t$, we have $\out= \sum_{i\in[d]}f^t(\inp_i)+\nu_{\ell}$ for some $\ell\in[\Sk]$, and, thus, $|\out- \sum_{i\in[d]}f^t(\inp_i)|\leq\alpha/8<\alpha$.

For $\delta=0$, we have
$\alpha= (8/\epsilon_1)\ln(2T/\beta)
=O(\sqrt{\epsilon^{-1}K\ln (T/\beta)} \ln(T/\beta)+\epsilon^{-1}\ln(T/\beta)).
$
       Plugging in $\Sk=\left(\frac{K\epsilon}{36\sqrt{\ln(1/\delta)}\ln(T/\beta)}\right)^{2/3}+1$ yields the final bound for $\delta>0$.
\end{proof}
To finish the proof of Theorem~\ref{thm:upperbound}, note that if $\epsilon^{-1}\ln(T/\beta)>\sqrt{\epsilon^{-1}K\ln(T/\beta)}$, then $\sqrt{\epsilon^{-1}K\ln(T/\beta)}>K$,
which can be seen by multiplying both sides of the inequality with $\sqrt{K}/\sqrt{\epsilon^{-1}\ln(T/\beta)}$.
Thus the upper bound $\min(d,K,\sqrt{\epsilon^{-1}K\ln(T/\beta)})$ holds for $\delta=0$.

Also, if $\epsilon^{-1}\sqrt{\ln(1/\delta)}\ln(T/\beta)>\left(\epsilon^{-2}K\ln(1/\delta)\ln^2(T/\beta)\right)^{1/3}$,
then
$\epsilon^{-1}\sqrt{\ln(1/\delta)}\ln(T/\beta)>K$,
which can be seen by first cubing the inequality and then dividing by $\epsilon^{-2}\ln(1/\delta)\ln^2(T/\beta)$.
Thus, for $\delta > 0$, the upper bound of $\min(d,K, \left(\epsilon^{-2}K\ln(1/\delta)\ln^2(T/\beta)\right)^{1/3})$ holds.
\end{proof}

 \subsection{Generalizations}

 We now argue about Theorem~\ref{thm:generalization}. Let $Q$ be a real-valued function on input streams from $\{-1,0,1\}$ and let $Q^t=Q(x_1,\dots,x_t)$. Further, let $Q$ be such that 1.) for any $x$ and $y$ which are neighboring, we have $|Q^t(x)-Q^{t}(y)|\leq 1$ for all time steps $t$, and 2.) $\sum_{t=1}^T|Q^t(x)-Q^{t-1}(x)|\leq K$. The first bound from Theorem~\ref{thm:generalization} is achieved by an algorithm that never updates the output, and the third bounds for $\epsilon$ and $(\epsilon,\delta)$-differential privacy are obtained by the Laplace and Gaussian mechanisms, respectively. The second bound for both $\epsilon$ and $(\epsilon,\delta)$-differential privacy is obtained by Algorithm~\ref{alg:monitoring} by setting $Q=Q^t(x)$ at every time step $t$. The proofs follow by exchanging $\sum_{i\in[d]} f^t(x_i)$ by $Q^t(x)$ in the proofs of Lemma~\ref{lem:monitoring_private} and \ref{lem:monitoring_accurate}.%

\section{A Connection between the General Model under Event-Level Privacy and the ``Likes"-Model under Item-Level Privacy}
Our bounds from Theorems~\ref{thm:upperbound_intro}, \ref{thm:intro_lower_eps}, and \ref{thm:lowerbound_epsdel_intro} as well as the bounds from~\cite{jain2023counting} imply that under \emph{item-level privacy}, the \simplemodel\ and the \generalmodel\ are roughly equally hard: all upper bounds hold for the \generalmodel\ and all lower bounds hold for the \simplemodel, and the bounds are tight up to a $\log T$ factor.
However, under \emph{event-level privacy}, the \simplemodel\ is significantly easier than the general model: It can be solved via continual counting on the difference sequence of the true output, which gives error polylogarithmic in $\log T$.
This is possible because for event-level privacy in the \simplemodel, the difference sequence of the output (i.e., the difference between the true output value of the current and the preceding time step) has $\ell_{\infty}$-sensitivity 1 for event-level privacy, but for item-level privacy, the sensitivity can be as large as $T$.

In the \generalmodel, there are no better upper bounds known for event-level differential privacy than for item-level differential privacy, and the upper and lower bounds from~\cite{jain2023counting} for $(\epsilon,\delta)$-differential privacy for the event-level setting in the \generalmodel\ %
leave a polynomial (in $T$) gap, in the case where the maximum flippancy $w_x \in (T^{1/2},T^{2/3})$: %
In that case, ignoring polynomial factors in $\epsilon^{-1}$, $\log(1/\delta)$, and $\log T$, the lower bound of~\cite{jain2023counting} is $\Omega(T^{1/4})$, while their algorithm gives an additive error of $O(T^{1/3})$.  Specifically, finding the best achievable error for \emph{event-level} privacy in the \generalmodel\ is explicitly posed as an open question in \cite{jain2023counting}.

We resolve this question for a large class of algorithms, called \emph{$\gamma$-output-determined} algorithms.
All known algorithms for this problem in \emph{any} model are 0-output-determined.
Specifically, we show that for $\gamma$-output-determined algorithms
our lower bounds and the lower bounds from \cite{jain2023counting} for \emph{item-level} privacy in the \emph{\simplemodel} basically carry over to \emph{event-level} privacy in the \emph{\generalmodel.}
It follows that our algorithm and the algorithm from \cite{jain2023counting} for event-level privacy in the \generalmodel\ are tight up to a factor that is linear in $\log T$ \emph{within the class of output-determined algorithms}.
Note that our reduction works both for the $\epsilon$-differential privacy as well as for $(\epsilon,\delta)$-differential privacy and we give the corresponding lower bounds in Theorems~\ref{thm:lower_eps} and~\ref{thm:lowerbound_epsdel}.
In the following, we denote by $\countdist(x)$ the stream of true answers to the $\countdist$ problem on stream $x$.
   \begin{definition}
   Let $\gamma \ge 0$.
        An algorithm $\Alg$ for the \countdist\ problem is said to be $\gamma$-\emph{\outdet}, if for all inputs $x$ and $y$ such that $\countdist(x)=\countdist(y)$ and any $S\in \mathrm{range}(\Alg)$ we have:
        \begin{align*}
            \Pr(\Alg(x)\in S) \le \Pr(\Alg(y)\in S) + \gamma
        \end{align*}
    \end{definition}

\begin{theorem}\label{thm:event-level}
Let $\epsilon > 0, \delta\ge 0$ and $\gamma \ge 0$.
Let $\Alg_1$ be an event-level, $(\epsilon,\delta)$-differentially private, $\gamma$-\outdet\ algorithm for \countdist\ that works in the \generalmodel\ and has error at most $\alpha$ for streams of length $T+1$ with probability $1-\beta$. Then there exists an item-level, $(2\epsilon,(1+e^{\epsilon})\delta + e^{\epsilon} \gamma)$-differentially private algorithm $\Alg_2$ for \countdist\ that works in the \simplemodel, and has error at most $\alpha$ for streams of length $T$
with probability $1-\beta$.
\end{theorem}
\begin{proof}
We describe algorithm $\Alg_2$, that is item-level $(2\epsilon,(1+e^{\epsilon})\delta+e^{\epsilon} \gamma)$-dp in the \simplemodel, derived from a $\gamma$-\outdet\ algorithm $\Alg_1$ which is event-level, $(\epsilon,\delta)$-dp in the \generalmodel: Let $x$ be an input for \countdist\ in the \simplemodel\ of length $T$, i.e., $x$ is such that $\sum_{t'\leq t}x_i^{t'}$ can only take the values $0$ or $1$, for any $i\in[d]$ and $t\in[T]$. %
 Let $x_0=0^d x$, i.e., we attach a $d$-dimensional all-zero vector before $x$, and define $(\Alg_2(x))^t=(\Alg_1(x_0))^{t+1}$ for all $t\in[T]$ (note that $\Alg_1$ can take inputs from the \simplemodel). We now show that $\Alg_2$ is item-level $(2\epsilon,(1+e^{\epsilon})\delta+e^{\epsilon} \gamma)$-differentially private. Let $x$ and $y$ be two item-level neighbouring inputs in the \simplemodel. That is, there exists an item $i$ such that the streams $x_i$ and $y_i$ may be completely different, while $x_j=y_j$ for all $j\neq i$. Additionally, since we are in the \simplemodel, for any time step $t$, $\sum_{t'\leq t} x_i^{t'}\in\{0,1\}$ and $\sum_{t'\leq t} y_i^{t'}\in\{0,1\}$.

Next, we define input streams $z$ and $w$ in the \generalmodel\ where $\countdist(z)=\countdist(w)$, $z$ is event-level neighbouring to $x_0$, and $w$ is event-level neighbouring to $y_0$. Since $\Alg_1$ is event-level $(\epsilon,\delta)$-dp and works for the \generalmodel, we then have for any $S\in \mathrm{range}(\Alg_2)$
\begin{align*}
\Pr[\Alg_2(x)\in S]
&=\Pr[(\Alg_1(x_0))_{t=2}^{T+1}\in S]
\leq e^{\epsilon}\Pr[(\Alg_1(z))_{t=2}^{T+1}\in S]+\delta \\
&\le e^{\epsilon}\Pr[(\Alg_1(w))_{t=2}^{T+1}\in S]+\delta +\gamma \\
&\leq e^{2\epsilon}\Pr[(\Alg_1(y_0))_{t=2}^{T+1}\in S]+(1+e^{\epsilon})\delta + e^{\epsilon} \gamma \\
&= e^{2\epsilon}\Pr[\Alg_2(y)\in S]+(1+e^{\epsilon})\delta + e^{\epsilon} \gamma,
\end{align*}
where the second inequality holds as $\Alg_1$ is $\gamma$-output-determined.

To define such $z$ and $w$, let $-e_i$ be the vector such that $-e_i(j)=0$ for all $j\neq i$ and $-e_i(i)=-1$. Then $z=-e_i x$ and $w=-e_i y$. Note that $z$ and $w$ are valid input streams for the \generalmodel, while they are not valid for the \simplemodel. Clearly, $z$ is event-level neighbouring to $x_0$, and $w$ is event level neighbouring to $y$.
 Recall that $\countdist(z)^t=\sum_{j=1}^{\unisize}\indfun(\sum_{t'\leq t}z_j^t>0)$. Since $\sum_{t'\leq t} x_i^t\in\{0,1\}$ for all $t\in[T]$ we have $\sum_{t'\leq t} z_i^t\leq 0$ for all $t\in[T+1]$. By the same argument, we have $\sum_{t'\leq t} w_i^t\leq 0$ for all $t\in[T+1]$. Since $z$ and $w$ only differ in the $i$th coordinate, which never contributes to the $\countdist$ value as it is never 1, we have $\countdist(z)=\countdist(w)$.

We are left with analyzing the error of the two algorithms. For this, note that by definition of $x_0$, we have $\countdist(x_0)^{t+1}=\countdist(x)^{t}$. Thus, running $\Alg_2$ on $x$ gives the same error as running $\Alg_1$ on $x_0$.
\end{proof}
In particular, for any \outdet\ algorithm, Theorem~\ref{thm:event-level} implies that all lower bounds on the error for the \countdist\ problem under \emph{item-level} differential privacy which hold for the\emph{\simplemodel}\ (and thus,  all lower bounds for \countdist\ under item-level differential privacy shown in this paper in Theorem~\ref{thm:lowerbound_epsdel} and in \cite{jain2023counting}), carry over to \emph{event-level} differential privacy in the \emph{\generalmodel}. This means that if there is an algorithm achieving a better error than the bounds stated in Theorem~\ref{thm:lowerbound_epsdel} and in \cite{jain2023counting} for event-level differential privacy in the \generalmodel, it cannot be $\gamma$-\outdet\ for $\gamma = O(\delta)$, i.e., it must be such that it does not \emph{only} depend on the number of distinct elements at any given time step.

\section{Item-Level Lower Bounds in the ``Likes"-Model}\label{sec:lowerbound}

In the following we show lower bounds for solving \countdist\ under item-level differential privacy, and in the \simplemodel. The lower bounds also apply to the \generalmodel. In \cref{sec:upperbound}, we showed a complementing upper bound which holds in the \generalmodel, even if $K$ is unknown to the algorithm.%
\begin{theorem}
\label{thm:lower_eps}
    Let  $d$ and $T > 4$ be non-negative integers and let $\epsilon > 0$.
    \begin{enumerate}
        \item Let $L\ge 8$ be a non-negative integer such that $L\le dT$. There exists an input stream $x$ of $d$-dimensional vectors from $\{-1,0,1\}^d$, which is valid in the \simplemodel\ with multiple updates per time step, with length $T$ and flippancy $K$ with
        $\min(3L/8, T/4-1) \le K \le \min(L, dT/4)$ such that any $\epsilon$-differentially private algorithm to the \countdist\ problem with item-level privacy with error at most $\alpha$ at all time steps with probability at least $2/3$ must satisfy \[\alpha=\Omega(\min(d,L,\epsilon^{-1}T, \sqrt{\epsilon^{-1}L\max(\ln (T/L) ,1)}))=\Omega(\min(d,K,\epsilon^{-1}T, \sqrt{\epsilon^{-1}K\max(\ln (T/K) ,1)})).\]
        \item Let $L \ge 8$ be a non-negative integer such that $L\le T$. There exists an input stream $x$ of $d$-dimensional vectors from $\{-1,0,1\}^d$,  which is valid in the \simplemodel\ with multiple updates per time step, with length $T$, flippancy $K$ with $L/16 \le K \le \min(L,T/4)$, and with $||x^t||_1=1$ for all $t$ (i.e., each update modifies at most one item) such that any $\epsilon$-differentially private algorithm to the \countdist\ problem with item-level privacy with error at most $\alpha$ at all time steps with probability at least $2/3$ must satisfy \[\alpha=\Omega(\min(d,K,\sqrt{\epsilon^{-1}K\ln (T/K)}).\]
    \end{enumerate}
\end{theorem}

\begin{proof}

Let $d$, $T$, and $L$ be as given in the theorem statement.
Assume there is an $\epsilon$-differentially private algorithm {${\mathcal A}$} for the \countdist\ problem with error at most $\alpha$ at all time steps with probability at least 2/3. If $\alpha>d/2$, then the error is $\Omega(d)$. Also, if $\alpha > L/8$, then $\alpha = \Omega(L)$.  Thus, in the following, we consider the case $\alpha\leq d/2$ and $\alpha \le L/8$.
Defining $m=\lfloor 2\alpha \rfloor$, it follows that $m \leq \min(d,L/8)$.

\paragraph*{Singleton updates}
We first find $T' \le T$ and $L' \le L$ such that $4m$ divides $T'$ and $m$ divides $L'$. If this is not the case for $T$ and $L$, then
pick parameters $T'$ and $L'$ such that (i) $4m$ divides $T'$ and $m$ divides $L'$, (ii) $\Delta = T-T' \le 4m < L/2 \le T/2$  (i.e. $T' = \Theta(T)$) and (iii) $0 \le L - \Delta - L' \le m$. This implies that $L' \ge 7L/8 - \Delta \ge 3L/8$. Thus, as $L \le T$, then
 $0\le L - \Delta - L' =  L- (T-T') - L' = T' - L' - (T-L) \le T'-L'$, i.e.,
 $L'\le T'$.

We use $T'$ and $L'$  in the proof below to construct a sequence of length $T'$ fulfilling the statements of the theorem. To complete the proof of the theorem, we
append to the sequence $T-T'$ many all-zero vectors to guarantee that the stream has length $T$. Note that appending to the sequence ``blank'' operation will not invalidate the statements of the theorem.

We now construct a set of input sequences of length $T'$ with flippancy $K:=\min(L', T'/4)$ and use them to prove a lower bound for $\alpha$ of $\Omega(\min(K \ln(T'/K), \sqrt{\epsilon^{-1}K \ln(T'/K)}))$. Combined with the above case distinctions giving lower bounds on $\alpha$ of $\Omega(d)$, and $\Omega(L)$, the fact that  $K =\Theta(L)$ and that $T' = \Theta(T)$, this implies that $\alpha = \Omega(\min(d,K, \sqrt{\epsilon^{-1}K(\ln(T/K)+1)})$.

Let $k := \min(L',T'/4)/m$ be a positive integer.
Partition the timeline into $T'/m$ blocks of length $m$, namely $B_1=[1,m]$, $B_2=[m+1,2m]$, $\dots$.
Now, for any subset of blocks $J=(j_1,\dots,j_k)$ with $1\leq j_1<j_2<\dots < j_k\leq T'/m$, define an input sequence $\inp(J)$ such that for any item $i\in[m]$
we insert element $i$ in the $i$th time step of every odd block of $J$ (i.e.~the first, third, ... block in $J$), and delete it again at the $i$th position of every even block of $J$ (i.e.~the second, fourth, ... block in $J$).
More formally,
for any item $i\in[m]$, set $\inp(J)_i^t=1$ for all $t=B_{j_{2p-1}}[i]=(j_{2p-1}-1)m+i$, $p=1\dots,\lceil k/2 \rceil$, and set $\inp(J)_i^t=-1$ for all $t=B_{j_{2p}}=(j_{2p}-1)m+i$, $p=1\dots,\lceil k/2 \rceil$.   %
In all other time steps $t$, no updates are performed, i.e., $\inp(J)^t$ is an all-zero vector. Thus, for every $i\in[m]$, we have $f^t(x_i)=1$ for all time steps $t\in[j_{2p-1}m,(j_{2p}-1)m]$,
for all $p\leq \lceil k/2\rceil$, and $f^t(x_i)=0$ for all time steps $t\in[j_{2p}m,(j_{2p+1}-1)m]$. For any item $m < i \le d$, we have $f^t(x_i)=0$ for all $t\in[T']$. Furthermore, items $i$ with $i >m$ (if they exist) are never inserted or deleted.  In total, there are $k=\min(L',T'/4)/m$ updates per item $i\in[m]$, thus exactly $K$ updates in total, and, hence, the total flippancy is $K = \min(L', T'/4)$.
If $K = L'$, then $L \ge K \ge 3L/8$. If $K = T'/4$, then  $L'\le T'$ implies that $L\ge L' \ge K = T'/4 \ge L'/4\ge 3L/32 \ge L/16$. Thus in either case $K = \Theta(L')$. Furthermore $K\le T'/4 \le T/4$.

Now let $E_J$, for  $J=(j_1,\dots,j_k)$ with $1\leq j_1<j_2<\dots < j_k\leq T'/m$, be the set of output sequences where {${\mathcal A}$} outputs (i) a value of $m/2$ or larger for all time steps $t\in[j_{2p-1}m,(j_{2p}-1)m]$ with $1\leq p\leq \lceil k/2\rceil$, and (ii) smaller than $m/2$ for all time steps $t$ such that (a) $t < j_1 m$ or (b) $t \in[j_{2p}m,(j_{2p+1}-1)m]$ for some $0\leq p< \lceil k/2\rceil $ or (c) $t \ge j_{k}m$. %
Note that for an input sequence  $x(J)$ every output sequence where {${\mathcal A}$} has additive error smaller than $\alpha = m/2$ must belong to $E_J$. As the algorithm is correct with probability at least $2/3$,
 $\Pr[{{\mathcal A}}(x(J))\in E_J]\geq 2/3$.

Two input sequences are neighboring if they differ
in the data of at most one item for item-level differential privacy. As two input sequences $x(I)$ and $x(J)$ with $I \ne J$ differ in the data of at most $m$ items, it follows
by group privacy that $\Pr[{{\mathcal A}}(x(J))\in E_I]\geq e^{-m\epsilon}2/3$ for any $J=(j_1,\dots,j_k)$ with $1\leq j_1<j_2<\dots < j_k\leq T'/m$ and $I=(i_1,\dots,i_k)$ with $1\leq i_1<i_2<\dots < i_k\leq T'/m$.
Also note that the  set of output sequences $E_J$ for distinct $J=(j_1,\dots,j_k)$ are disjoint, since for each multiple of $m$ (i.e., the end of a block), it is clearly defined whether the output is at least $m/2$ or smaller than $m/2$, and as such the values $j_1,\dots,j_k$ can be uniquely recovered.
Thus, there are $\binom{T'/m}{k}$ disjoint events $E_J$ and the sum over all $J$ of the probabilities that the algorithm with input $x(I)$  outputs an  event $E_J$ is  at most 1.
More formally, we have:
\begin{align*}
1\geq  \binom{T'/m}{k} e^{-m\epsilon}2/3\geq \frac{(T'/m)^{k}}{(k)^{k}}e^{-m\epsilon}2/3. %
= \frac{T'^{(K/m)}}{K^{K/m}}e^{-m\epsilon}2/3
\end{align*}
where the last equality is since $k=K/m$.
This gives
\[
m^2 +\epsilon^{-1} m \ln(3/2) \geq \epsilon^{-1}K\ln(T'/K)
\]
which implies
\[
m = \Omega(\min(K \ln(T'/K),
\sqrt{\epsilon^{-1}K\ln(T'/K)}).
\]
Note that since $T' \ge 4K$, $\ln(T'/K)\geq \ln(4)>1$. This completes the proof.

\paragraph*{Multiple updates}
We first find $T' \le T$ and $L' \le L$ such that $4$ divides $T'$ and $m$ divides $L'$. If this is not the case for $T$ and $L$, then
pick parameters $T'$ and $L'$ such that (i) $4$ divides  $T'$ and $m$ divides $L'$, (ii) $\Delta = T-T' \le 4$ (i.e. $T' = \Theta(T)$) and (iii) $\Delta m \le L - L' \le (\Delta + 1) m$. This implies that $L' \ge L - (\Delta+1)m \ge L - 5m \ge 3L/8$.

We use $T'$ and $L'$  in the proof below to construct a sequence of length $T'$ fulfilling the statements of the theorem. To complete the proof of the theorem, we
append to the sequence $T-T'$ many all-zero vectors to guarantee that the stream has length $T$. Note that appending to the sequence ``blank'' operation will not invalidate the statements of the theorem.

The idea is similar to above, only we do not define blocks, but directly choose $k:=\min(L'/m,T'/4)$ time steps in which all items in $[m]$ are updated. Thus the flippancy $K$ will equal $mk$.
More precisely, we construct the following set of input sequences. For any $I=(t_1,\dots,t_k)$ with $1\leq t_1<t_2<\dots<t_k\leq T'$, we define an input sequence $\inp(I)$ as follows: For any item $i\in[m]$, set $\inp(I)_i^{t_{j}}=1$ for all odd $j$, and $\inp(I)_i^{t_{j}}=-1$ for all even $j$. All other coordinates are set to $0$. In total, there are $k$ updates per item in $[m]$, thus, exactly $K$ updates in total, i.e., the total flippancy equals $K = \min(L', mT'/4)$.
This implies that $\min(3L/8, T/4-1) \le K \le \min(L, dT/4)$.

Now, let $E_I$, for $I=(t_1,\dots,t_k)$ with $1\leq t_1<t_2<\dots<t_k\leq T'$, be the set of output sequences with a value of $m/2$ or larger at all time steps $t\in[t_{2p-1},t_{2p})$ for some $1\leq p\leq \lceil k/2\rceil$, and a value smaller than $m/2$ at all time steps $t$ where (a) $t\leq t_1$ or (b) $t\in[t_{2p},t_{2p+1})$ for some $0\leq p<\lceil k/2\rceil$. Note that for input sequence $\inp(I)$ every output sequence where $\Alg$ has an additive error smaller than $m/2$ must be in $E_I$. As the algorithm is correct with probability at least 2/3, $\Pr[\Alg(\inp(I))\in E_I]\geq 2/3$.
As two input sequences $x(I)$ and $x(J)$ with $I \ne J = (j_1, \dots , j_k)$
 with $1\leq j_1<j_2<\dots < j_k\leq T'$
differ in the data of at most $m$ items, it follows
by group privacy that $\Pr[{{\mathcal A}}(x(I))\in E_J]\geq e^{-m\epsilon}2/3$ for any such $J$.

Let $J=(j_1,\dots,j_k)$ with $1\leq j_1<j_2<\dots<j_k\leq T'$.
Note that the  events $E_I$ and $E_J$ for any $I \ne J$ are disjoint, since in the event $E_I$ it is clearly defined for every time step whether the output is at least $m/2$ or smaller than $m/2$, and from that the set $I$ can be uniquely recovered.
Thus, there are $\binom{T'}{k}$ disjoint events $E_J$ and the probability that with input $x(I)$ the algorithm  outputs any one of them is at most 1.
Thus we have
\begin{align}\label{inequ:1}
    1
    \geq \binom{T'}{k}e^{-m\epsilon}2/3
    \geq \frac{T'^k}{k^k}e^{-m\epsilon}2/3
    = \frac{T'^{K/m}}{{(K/m)}^{K/m}}e^{-m\epsilon}2/3
\end{align}
where the last equality is since $k=K/m$.

Next we consider two cases, the first one resulting in two different lower bounds on $m$ and the second one giving a third lower bound on $m$. The combination of these three lower bounds then gives the claimed bound above of
$$\alpha = m/2 = \Omega(\min(\epsilon^{-1}T', \sqrt{\epsilon^{-1}K \max(\ln(T'/K), 1)},  K \max(\ln(T'/K),1)) ))$$

\noindent
\textbf{Case 1:} $L'< mT'/4$.
In this case $K = L'$ and we have
\begin{align*}
    m^2\epsilon + m \ln(3/2)\geq K\ln(T'm/K)\geq K\max(\ln(T'/K),1)
\end{align*}
where the last inequality holds since $K \le mT'/4$, i.e., $\ln( T'm/K) \ge \ln(4) > 1$. Hence
\begin{align*}
    m=\Omega(\min(\sqrt{\epsilon^{-1}K \max(\ln(T'/K),1)}, K \max(\ln(T'/K),1))).
\end{align*}
As $K = L' = \Theta(L)$ it follows that
\begin{align*}
m=\Omega(\min(\sqrt{\epsilon^{-1}L\max(\ln(T'/L),1)}, L \max(\ln(T'/L),1))).
\end{align*}

\noindent
\textbf{Case 2:} $L'\ge mT'/4$.
This implies that $K=mT'/4$ and, thus, that there are updates in at least $T'/4$ many time steps. In this case Inequality~\ref{inequ:1} can  be reformulated as follows:
\begin{align*}
    1\geq \frac{T'^{K/m}}{{K/m}^{K/m}}e^{-m\epsilon}2/3 =
    4^{T'/4} e^{-m\epsilon}2/3 = e^{\ln(4) T'/4 - m\epsilon}2/3,
\end{align*}
which implies that Inequality~\ref{inequ:1} is satisfied for $m = \Omega(\epsilon^{-1} T')$.

These two cases show that  $\alpha = \Omega(\min(\epsilon^{-1}T', \sqrt{\epsilon^{-1}L \max(\ln(T'/L), 1)},  L \max(\ln(T'/L),1)) )$ for the above input sequence.
Combined with the above lower bounds on $\alpha$ of $\Omega(\min(d, L))$ and the fact that  %
$T'= \Theta(T)$, it follows that $\alpha =
\Omega(\min(d, L, \epsilon^{-1}T, \sqrt{\epsilon^{-1}L \max(\ln(T/L), 1)}))$.
\end{proof}
\section{Lower Bounds for Approximate Differential Privacy}

In this section, we adapt the lower bounds from \cite{jain2023counting} for item-level differential privacy to our parameter scheme.

\begin{theorem}\label{thm:lowerbound_epsdel}
Let $\epsilon,\delta\in(0,1]$.
\begin{enumerate}
    \item \label{thm:lowerbound_epsdel_general}Let $K, T$ be sufficiently large parameters. There exists a dimension $d$ and an input stream $x$ of $d$-dimensional vectors from $\{-1,0,1\}^d$ of length $T$ and with flippancy at most $K$ which is valid in the ``likes''-model, such that any item-level, $(\epsilon,\delta)$-differentially private algorithm to the \countdist\ problem with error at most $\alpha$ at all time steps with probability at least 0.99 must satisfy $\alpha=\Omega\left(\min\left(\frac{\sqrt{T}}{\epsilon\log T},\frac{(K\epsilon)^{1/3}}{\epsilon\log (K\epsilon)}\right)\right)$.
    \item \label{thm:lowerbound_epsdel_singleton}Let $K$ and $T$ be sufficiently large parameters satisfying $K\leq T$. There exists a dimension $d$ and
    an input stream $x$ of $d$-dimensional vectors from $\{-1,0,1\}^d$ of length $T$ and with flippancy at most $K$ which is valid in the ``likes''-model and satisfies $||x^t||_1=1$ for all $t$, such that any item-level, $(\epsilon,\delta)$-differentially private algorithm to the \countdist\ problem with error at most $\alpha$ at all time steps with probability at least 0.99 must satisfy $\alpha=\Omega\left(\frac{K^{1/3}}{\epsilon\log K}\right)$.
\end{enumerate}
\end{theorem}
The reduction in \cite{jain2023counting} is based on a lower bound for the 1-way marginals problem. In that problem, the data set $y$ is an table consisting of $n$ rows and $m$ columns, where every entry is in $\{0,1\}$. Two data sets $y$ and $y'$ are neighbouring if they differ in at most one row. The goal is to estimate the average column sums, i.e., the vector $(\sum_{i=1}^n y[i,j])_{j\in[m]}$. The following lower bound holds for estimating 1-way marginals under $(\epsilon,\delta)$-differential privacy:
\begin{lemma}[Bun, Ullman, and Vadhan~\cite{DBLP:journals/siamcomp/BunUV18}]\label{lem:marginals_lower}
Let $\epsilon\in (0,1]$, $\gamma\in(0,1)$, and $m,n\in\mathbb{N}$, and $\delta=o(1/n)$. Any algorithm which is $(\epsilon,\delta)$-differential private and has error at most $\gamma$ with probability at least 0.99 satisfies $n=\Omega\left(\frac{\sqrt{m}}{\gamma\epsilon\log m}\right)$.
\end{lemma}
\begin{proof}[Proof Sketch of Theorem~\ref{thm:lowerbound_epsdel}.]
We start by arguing about \cref{thm:lowerbound_epsdel_singleton}. For this case, our example stream is exactly the same as in \cite{jain2023counting}, given in Algorithm 5 in \cite{jain2023counting} (for a formulation using our slightly different notation see Algorithm~\ref{alg:red_marg}). They give a reduction from the 1-way marginals problem:  For any instance $\mathcal{I}$ of the 1-way marginals problem with $n$ rows and $m$ columns, there is an instance $C(\mathcal{I})$ of \countdist\ with $T=2mn$, such that if $\mathcal{I}$ and $\mathcal{I'}$ are neighbouring, then $C(\mathcal{I})$ and $C(\mathcal{I'})$ are item-neighbouring. Further, if we can solve $C(\mathcal{I})$ within error $\alpha$, we can solve $\mathcal{I}$ within error $\alpha/n$. It follows by Lemma~\ref{lem:marginals_lower} that $\alpha=\Omega\left(\min\left(\frac{\sqrt{m}}{\epsilon\log m},n\right)\right)$. In the instance they constructed, $d=n$, i.e. each row in the 1-way marginals problem gives an item in the \countdist\ problem. Further, the total flippancy $K$ can be as large as $2mn$ for worst case inputs. Thus, in order to apply the reduction, we need $2mn\leq K\leq T$. Given parameters $K\leq T$, we choose $m=K/(2n)$. The lower bound $\Omega\left(\min\left(\frac{\sqrt{m}}{\epsilon\log m},n\right)\right)$ translates to $\Omega\left(\min\left(\frac{\sqrt{K/(2n)}}{{\epsilon}\log(K/(2n))},n\right)\right)$. For $n= \frac{K^{1/3}}{2(\epsilon\log K)^{2/3}}$, we have
\begin{align*}
\frac{\sqrt{K/(2n)}}{{\epsilon}\log(K/(2n))}\geq\frac{K^{1/3}\epsilon^{1/3}\log^{1/3} K}{\epsilon\log (K^{1/2})}=\Omega\left(\frac{K^{1/3}\log^{1/3} K}{\epsilon^{2/3}\log K}\right)=\Omega(n).
\end{align*}
Thus, we get $\alpha=\Omega\left(\frac{K^{1/3}\log^{1/3} K}{\epsilon^{2/3}\log K}\right)$.

For~\cref{thm:lowerbound_epsdel_general}., where we allow general updates, we have to slightly modify the example in \cite{jain2023counting}: namely, in their Algorithm 5, we collapse every one of their vectors $z^{(j)}$, $j=1,\dots,m$, into vectors of length 2, one time step for all insertions corresponding to column $j$, and one time step for all deletions corresponding to column $j$. See Algorithm~\ref{alg:red_marg_new}. We then again get a reduction with the same properties as before, except that  $T=2n$ and $K$ can be as large as $2mn$. Now, the analysis from~\cite{jain2023counting} can be repeated with our $T$ taking the role of $w_x$ in~\cite{jain2023counting}, and our $K$ taking the role of $T$~in~\cite{jain2023counting}.
\end{proof}

\begin{algorithm}[tb]

\begin{algorithmic}[1]
\STATE{\bf Input:} Data Set $y[1],\dots,y[n]\in\{0,1\}^{n\times m}$ and blackbox access to a mechanism $M$ for \countdist
\STATE{\bf Output:} Estimates of marginals $b=(b[1],\dots,b[m])$
\FOR{$j=1,\dots,m$}
\FOR{$i=1,\dots,n$}
    \STATE Set $z^{(j)}[i]=e_i$\;
    \STATE Set $z^{(j)}[i+n]=-e_i$\;
    \ENDFOR\ENDFOR
    \STATE Run $M$ on $x\rightarrow z^{(1)}\circ z^{(2)}\circ \dots \circ z^{(m)}$ and record answer vector $r$\;
    \FOR{$j\in[m]$ do}
    \STATE $b[j]=r[(2j-1)n]/n$
    \ENDFOR
\STATE {\bf output} b
\end{algorithmic}
\caption{Algorithm 5 from \cite{jain2023counting}: Reduction from 1-way marginals to \countdist}
\label{alg:red_marg}
\end{algorithm}

\begin{algorithm}[tb]
\begin{algorithmic}[1]
\STATE {\bf Input:} Data Set $y[1],\dots,y[n]\in\{0,1\}^{n\times m}$ and blackbox access to a mechanism $M$ for \countdist
\STATE {\bf Output:} Estimates of marginals $b=(b[1],\dots,b[m])$
\FOR{$j=1,\dots,m$}
    \STATE Set $z^{(j)}[1]=y^T[j]$
   \STATE  Set $z^{(j)}[2]=-y^T[j]$
    \ENDFOR
    \STATE Run $M$ on $x\rightarrow z^{(1)}\circ z^{(2)}\circ \dots \circ z^{(m)}$ and record answer vector $r$\;
    \FOR{$j\in[m]$ do}
    \STATE $b[j]=r[(2j-1)]/n$
    \ENDFOR
\STATE {\bf output} b
\end{algorithmic}
\caption{ Reduction from 1-way marginals to \countdist\ for arbitrarily many updates per round}
\label{alg:red_marg_new}
\end{algorithm}

\paragraph*{Funding}

\begin{wrapfigure}{r}{0.15\textwidth}
\includegraphics[width=0.13\textwidth]{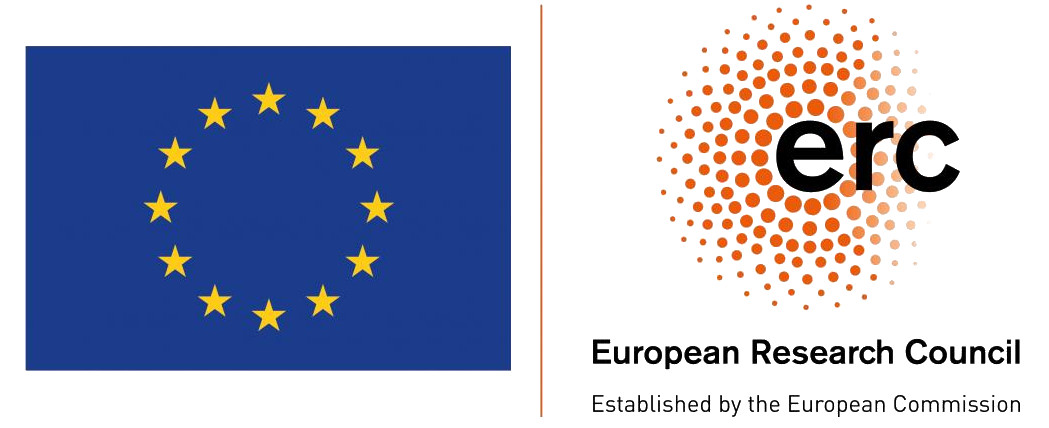}
\end{wrapfigure}

\textit{M. Henzinger}: This project has received funding from the European Research Council (ERC) under the European Union's Horizon 2020 research and innovation programme (MoDynStruct, No. 101019564) and the Austrian Science Fund (FWF) grant DOI 10.55776/Z422, grant DOI 10.55776/I5982, and grant DOI 10.55776/P33775 with additional funding from the netidee SCIENCE Stiftung, 2020–2024.

\textit{T. A. Steiner}: This work was supported by a research grant (VIL51463) from VILLUM FONDEN. %
\bibliographystyle{gamma}
\bibliography{References}
\appendix
\section{Unknown Total Flippancy}\label{upper_bound_unknownK}

\begin{algorithm}[tb]
\begin{algorithmic}[1]
\STATE {\bf Input:} Data Set $D=x^1, x^2,\dots$, initial counts $c_1,\dots,c_d$ (default 0), parameters $\epsilon$, $\delta$ and  $\beta$, $T$%
\STATE $t=1$\;
\STATE $K_1=2$\;
\FOR{$j=1,\dots,$}
   \STATE $\epsilon_j=6\epsilon/(\pi^2j^2)$\;
   \STATE $\delta_j=6\delta/(\pi^2j^2)$\;
   \STATE $\beta_j=6\beta/(\pi^2 j^2)$\;
    \IF{$\delta=0$}
    \STATE $\Skj=\sqrt{K_j\epsilon_j/(18\ln(T/\beta_j))}+1$
    \ENDIF

    \IF{$\delta>0$}
    \STATE $\Skj=\left(\frac{K_j\epsilon_j}{36 \sqrt{\ln(1/\delta_j)}\ln(T/\beta_j)}\right)^{2/3}+1$\ENDIF
    \STATE Run Algorithm~\ref{alg:monitoring} on $x^t,x^{t+1},\dots$, $c_1,\dots,c_d$, $\epsilon_j$, $\delta_j$, $\beta_j$, $T$, $\Skj$ until it aborts\; %
   \STATE  Let $t'$ be the last time step processed by Algorithm~\ref{alg:monitoring}\; %
    \STATE $t=t'+1$\;
   \STATE $j=j+1$\;
    \STATE $K_j=2^j$\;
\ENDFOR
\end{algorithmic}
\caption{\countdist, unknown $K$}
\label{alg:monitoring_unknownK}
\end{algorithm}

The algorithms from Section~\ref{sec:upperbound} can be easily extended to the case where the total flippancy $K$ is not known beforehand, at the cost of $\mathrm{polylog}(K)$ factors in the error bound, as shown by Algorithm~\ref{alg:monitoring_unknownK} and the lemmata below.
The fact that $K$ is not known causes no serious problem, as the algorithm repeatedly ``guesses'' $K$ and then runs the algorithm from earlier with the current guess.

\begin{lemma}
For any $0 < \epsilon < 1$ and $0 \le \delta < 1$,
Algorithm~\ref{alg:monitoring_unknownK} is $(\epsilon,\delta)$-differentially private.
\end{lemma}
\begin{proof}
    By Lemma~\ref{lem:monitoring_private}, the $j$th instance of Algorithm~\ref{alg:monitoring} is $(\epsilon_j,\delta_j)$-differentially private. Since $\sum_{j=1}^{\infty}\epsilon_j=\epsilon$ and $\sum_{j=1}^{\infty}\delta_j=\delta$, by Fact~\ref{fact:composition_theorem}, Algorithm~\ref{alg:monitoring_unknownK} is $(\epsilon,\delta)$-differentially  private.
\end{proof}
\begin{lemma}\label{lem:monitoring_unknownK}
For $\delta=0$, the error of Algorithm~\ref{alg:monitoring_unknownK} is at most \[O(\ln K\sqrt{\epsilon^{-1}K \ln(T\ln K /\beta)}+\epsilon^{-1}\ln^2 K \ln(T \ln K /\beta)).\] For $\delta>0$, the error of Algorithm~\ref{alg:monitoring_unknownK} is at most \[O((\epsilon^{-1}K\ln^2 K \ln(\ln K /\delta)\ln^2(T\ln K /\beta))^{1/3}+\epsilon^{-1}\ln^2 K\sqrt{\ln (\ln K /\delta)}\ln(T\ln K /\beta)).\]
\end{lemma}
\begin{proof}
    Let $j_l$ be the value of variable $j$ after the last element in the stream is processed. For any $j<j_l$, note that by Lemma~\ref{lem:monitoring_accurate}, with probability at least $1-\beta_j$, by the choice of $S_{K_j}$, the algorithm does not abort before having seen the entire stream if the total flippancy is at most $K_j$. Thus, when the algorithm aborts for some $j<j_l$, we know that the flippancy is at least $K_j$, %
    and the bound from Lemma~\ref{lem:monitoring_accurate} holds for the $j$th instance of Algorithm~\ref{alg:monitoring} with $\Skj$.

    Since the algorithm aborts for all $j<j_l$, we can conclude that the total flippancy of the stream processed by the $j$th run of Algorithm~\ref{alg:monitoring} is at least $K_j$. %
    Since $\sum_j \beta_j=\beta$, with probability at least $1-\beta$, (1) the total flippancy $K$ is at least $\sum_{j<j_l}K_j=2^{j_l}-1$, and (2) the bound from Lemma~\ref{lem:monitoring_accurate} holds for all instances of Algorithm~\ref{alg:monitoring} (with their respective parameters). It follows (a) that $K\geq K_{j_l}-1\geq K_j$ for all $j< j_l$ and (b) $j_l=O(\ln K)$. The maximum error over the stream is the maximum error of any instance of Algorithm~\ref{alg:monitoring}. %
    Since $K_j=O(K)$, $\epsilon_{j_l}\leq \epsilon_j$ and $\delta_{j_l}\leq \delta_j$ for all $j \le j_l$, the final bound is now obtained by plugging $K$, $\epsilon_{j_l} = \Theta(\epsilon/j^2)$ for $\epsilon$, $\delta_{j_l} = \Theta(\delta/j^2)$ for $\delta$, and $\beta_{j_l} = \Theta(\beta/j^2)$ for $\beta$ into the bound from Lemma~\ref{lem:monitoring_accurate}, and upper bounding $j^2$ by $\log^2 K$.
\end{proof}

One can also obtain the minimum of the bound from \cref{lem:monitoring_unknownK} and $\min(K, T, d)$ at the cost of an additive $\epsilon\ln^2K\ln(\ln K/\beta)$ factor with a slightly more involved algorithm, which involves choosing to either not update the output or abort if there is a trivial algorithm which performs better for the current estimate of $K$. If we knew the value of $K$ beforehand, we could choose the best algorithm upfront. Not knowing the value of $K$ makes it slightly more complicated.
In the following, we show how to obtain this bound. The full algorithm is given in Algorithm~\ref{alg:monitoring_unknownKminbound}.
\begin{theorem}\label{thm:upperbound_unknownK}
 Let $d$ and $T$ be non-zero integers, let $\beta>0$. Let $T$ be a known upper bound on the number of time steps. Then there exists
   \begin{enumerate}
       \item an item-level $\epsilon$-differentially private algorithm for  the \countdist\ problem in the general model with error at most \[O(\min(d,K,\ln K\sqrt{\epsilon^{-1}K\ln (T/\beta)},\epsilon^{-1}T\log(T/\beta))+\epsilon^{-1}\ln^2 K \ln(\ln K/\beta))\] at all time steps with probability at least $1-\beta$, for any $\epsilon>0$, where $K$ is the total flippancy of the input unknown to the algorithm. %
       \item an item-level $(\epsilon,\delta)$-differentially private algorithm for  the \countdist\ problem in the general model with error at most
       \[O(\min(d,K,(\epsilon^{-2}K\ln^2 K\ln(1/\delta)\ln^2(T/\beta))^{1/3},\epsilon^{-1}\sqrt{T\ln(1/\delta)\log(T/\beta)}+\epsilon^{-1}\ln^2 K \ln(\ln K/\beta)))\]
       at all time steps with probability at least $1-\beta$, for any $0<\delta<1$ and $0<\epsilon<1$, where $K$ is the total flippancy of the input unknown to the algorithm. %
   \end{enumerate}
\end{theorem}
\begin{proof}

    Privacy follows since Algorithm~\ref{alg:monitoring_unknownKminbound} is a composition of A) a post-processing of Algorithm~\ref{alg:monitoring_unknownK} with parameters $\epsilon/2$ and $\delta$ and B) a sequence of Laplace mechanisms such that the $j$th Laplace mechanism is $\epsilon_j$-differentially private, and $\sum_{j=1}^{\infty} \epsilon_j=\epsilon/2$.

\begin{algorithm}[!ht]
\begin{algorithmic}[1]

\STATE {\bf Input:} Data Set $D=x^1, x^2,\dots$, initial counts $c_1,\dots,c_d$ (default 0), parameters $\epsilon$, $\delta$ and  $\beta$, $T$%
\STATE $t=1$\;
\STATE $K_1=2$\;
\FOR{$j=1,\dots,$}
    \STATE $\epsilon_j=12\epsilon/(\pi^2j^2)$\;
    \STATE $\delta_j=6\delta/(\pi^2j^2)$\;
    \STATE $\beta_j=12\beta/(\pi^2 j^2)$\;
    \IF{$\delta=0$}
    \STATE $\Skj=\sqrt{K_j\epsilon_j/(18\ln(T/\beta_j))}+1$\;
    \STATE $B_j=\sqrt{\epsilon^{-1}K\ln (T/\beta)}$\;
   \STATE  $\err_T=\epsilon^{-1}T\log(T/\beta)$\ENDIF
    \IF{$\delta>0$}
    \STATE $\Sk=\left(\frac{K_j\epsilon_j}{36 \sqrt{\ln(1/\delta_j)}\ln(T/\beta_j)}\right)^{2/3}+1$\;
    \STATE $B_j=\left(\epsilon_j^{-2}K_j\ln(1/\delta_j)\ln^2(T/\beta_j)\right)^{1/3}+\epsilon_j^{-1}\sqrt{\ln(1/\delta_j)}\ln(T/\beta_j)$\;
    \STATE $\err_T=\epsilon^{-1}\sqrt{T\ln(1/\delta)\log(T/\beta)})$\ENDIF
        \IF{$\min(K_j,B_j)>\min(d,\err_T)$}
        \IF{$\min(d,\err_T)=d$}
   \STATE {\bf Abort} and switch to the trivial algorithm outputting 0 at all time steps;
    \ENDIF
     \IF{$\min(d,\err_T)=\err_T$}
    \STATE {\bf Abort} and switch to the Laplace mechanism with sensitivity bound $\Delta_1=T$, if $\delta=0$, and to the Gaussian mechanism with sensitivity bound $\Delta_2=\sqrt{T}$, else;
    \ENDIF\ENDIF
    \IF{$K_j\geq B_j$}
    \STATE Run Algorithm~\ref{alg:monitoring} on $x^t,x^{t+1},\dots$, $c_1,\dots,c_d$, $\epsilon_j$, $\delta_j$, $\beta_j$, $T$, $\Skj$ until it aborts\ENDIF
    \IF{$K_j<B_j$}
    \STATE Run Algorithm~\ref{alg:monitoring}$^*$ on $x^t,x^{t+1},\dots$, $c_1,\dots,c_d$, $\epsilon_j$, $\delta_j$, $\beta_j$, $T$, $\Skj$ until it aborts, where Algorithm~\ref{alg:monitoring}$^*$ is equal to Algorithm~\ref{alg:monitoring} without ever updating $\out$\ENDIF
   \STATE Let $t'$ be the last time step processed by Algorithm~\ref{alg:monitoring}\;
   \STATE $\out=\sum_{i=1}^d f^{t'}(x_i)+\Lap(1/\epsilon_j)$\;%
   \STATE $t=t'+1$\;
   \STATE $j=j+1$\;
   \STATE $K_j=2^j$\;
\ENDFOR
\end{algorithmic}

\caption{\countdist, unknown $K$, all bounds}
\label{alg:monitoring_unknownKminbound}

\end{algorithm}

    For accuracy, first assume that we do not abort before we have seen the entire stream. The updating parameters $t$, $j$ and $K_j$ are exactly the same as in Algorithm~\ref{alg:monitoring_unknownK}. Thus, we have $j\leq \log^2 K$ and $K_j=O(K)$ for all $j$ used in the algorithm. Next, we condition on 1) the accuracy bound from Lemma~\ref{lem:monitoring_accurate} holding for all runs of Algorithm~\ref{alg:monitoring_unknownK} resp. Algorithm~\ref{alg:monitoring_unknownK}$^{*}$ for their respective parameters, and 2) the Laplace noise $\mu_j\sim\Lap(1/\epsilon_j)$ in line ... satisfying $|\mu_j|\leq \epsilon_j^{-1}\log(1/\beta_j)$ for any $j$. By Lemma~\ref{lem:monitoring_accurate}, 1) is true with probability at least $1-\sum_j \beta_j\geq 1-\beta/2$. By the Laplace tailbound Fact~\ref{fact:laplace_tailbound}, 2) is true with probability at least $1-\sum_j \beta_j\geq 1-\beta/2$. Thus, 1) and 2) hold together with probability at least $1-\beta$.

    Now, consider first the case where $\delta=0$. For any $j$ such that $K_j>B_j$, we have that the error of the run of Algorithm~\ref{alg:monitoring} in the $j$th round is bounded by $O(\sqrt{\epsilon_j^{-1}K_j\ln (T/\beta_j)}+\epsilon_j^{-1}\ln(T/\beta_j))$ by our condition. Note that since $K_j>B_j=\sqrt{\epsilon_j^{-1}K_j\ln (T/\beta_j)}$, we have that $\epsilon_j^{-1}\ln(T/\beta_j)<\sqrt{\epsilon_j^{-1}K_j\ln (T/\beta_j)}$, which can be seen by multiplying both sides of the inequality with $\sqrt{\epsilon_j^{-1}K_j^{-1}\ln(T/\beta_j)}$. Thus, for any $j$ such that $K_j>B_j$, the error is \[O(\sqrt{\epsilon_j^{-1}K_j\ln (T/\beta_j)}=O(\min(K_j,\sqrt{\epsilon_j^{-1}K_j\ln (T/\beta_j)})=O(\min(K,\ln K\sqrt{\epsilon^{-1}K\ln (T\ln K/\beta)}).\] For any $j$ such that $K_j\leq B_j$, we do not update the output until the end of round $j$. Let $t_{j-1}$ resp. $t_j$ be the time step where the $(j-1)$st resp. $j$th copy of Algorithm~\ref{alg:monitoring} or  Algorithm~\ref{alg:monitoring}$^*$ aborted. Note that by Lemma~\ref{lem:monitoring_accurate}, the flippancy in interval $[t_{j-1}, t_j)$ is at most $K_j$. Thus, $|\sum_{i=1}^d f^{t_{j-1}}(x_i)-\sum_{i=1}^d f^{t'}(x_i)|\leq K_j$, for all $t\in[t_{j-1}, t_j)$. Since the output at all those time steps is given by $\out=\sum_{i=1}^d f^{t_{j-1}}(x_i)+\mu_j$, where $|\mu_j|\leq \epsilon_j^{-1}\log(1/\beta_j)$. By triangle inequality, we have \begin{align*}|\sum_{i=1}^d f^{t_{j-1}}(x_i)-\out|&\leq K_j+\epsilon_j^{-1}\log(1/\beta_j)\\&=O(\min(K_j,\sqrt{\epsilon_j^{-1}K_j\ln (T/\beta_j)})+\epsilon_j^{-1}\log(1/\beta_j)\\&=O(\min(K,\ln K\sqrt{\epsilon^{-1}K\ln (T\ln K/\beta)})+\epsilon^{-1}\ln^2 K \ln(\ln K/\beta)).\end{align*}

    Next, consider the case where $\delta>0$. For any $j$ such that $K_j>B_j$, we have that the error of the run of Algorithm~\ref{alg:monitoring} in the $j$th round is bounded by
    \[O\left(\left(\epsilon_j^{-2}K_j\ln(1/\delta_j)\ln^2(T/\beta_j)\right)^{1/3}+\epsilon_j^{-1}\sqrt{\ln(1/\delta_j)}\ln(T/\beta_j)\right)\]
    by our conditioning. Note that if $\epsilon_j^{-1}\sqrt{\ln(1/\delta_j)}\ln(T/\beta_j)>\epsilon_j^{-2}K_j\ln(1/\delta_j)\ln^2(T/\beta_j)$, then this gives us that $\epsilon_j^{-1}\sqrt{\ln(1/\delta_j)}\ln(T/\beta_j)>K_j$, in contradiction to $K_j>B_j$.  Thus, for any $j$ such that $K_j>B_j$, the error is \begin{align*}O\left(\left(\epsilon_j^{-2}K_j\ln(1/\delta_j)\ln^2(T/\beta_j)\right)^{1/3}\right)&=O\left(\min(K_j,\left(\epsilon_j^{-2}K_j\ln(1/\delta_j)\ln^2(T/\beta_j)\right)^{1/3}\right)\\&=O\left(\min(K,\left(\ln^2 K\epsilon^{-2}K\ln(\ln K/\delta)\ln^2(T\ln K/\beta)\right)^{1/3}\right).\end{align*} For any $j$ such that $K_j\leq B_j$, we do not update the output until the end of round $j$. Let $t_{j-1}$ resp. $t_j$ be the time step where the $(j-1)$st resp. $j$th copy of Algorithm~\ref{alg:monitoring} or  Algorithm~\ref{alg:monitoring}$^*$ aborted. Note that by Lemma~\ref{lem:monitoring_accurate}, the flippancy in interval $[t_{j-1}, t_j)$ is at most $K_j$. Thus, $|\sum_{i=1}^d f^{t_{j-1}}(x_i)-\sum_{i=1}^d f^{t'}(x_i)|\leq K_j$, for all $t\in[t_{j-1}, t_j)$. Since the output at all those time steps is given by $\out=\sum_{i=1}^d f^{t_{j-1}}(x_i)+\mu_j$, where $|\mu_j|\leq \epsilon_j^{-1}\log(1/\beta_j)$. By triangle inequality, we have $|\sum_{i=1}^d f^{t_{j-1}}(x_i)-\out|\leq K_j+\epsilon_j^{-1}\log(1/\beta_j)$.

    Lastly, we argue about what happens if we abort and switch to one of the trivial algorithms. Note that this happens exactly when we reach a $j$ such that $\min(K_j,B_j)>\min(d,\err_T)$. Note that up to $j-1$, by the analysis before, we have that the error is bounded by $\min(K_{j-1},B_{j-1})+\epsilon_{j-1}^{-1}\log(1/\beta_{j-1})\leq \min(d,\err_T)+\epsilon_{j-1}^{-1}\log(1/\beta_{j-1})$, since the algorithm did not abort. After we abort, the error of the algorithm is $O(\min(d,\err_T))$. Thus, the algorithm at any time step is bounded by $O(\min(d,\err_T))+\ln^2K\epsilon^{-1}\log(1/\beta)$. Since we have that $K_j\leq K$ , we have for $\delta = 0$ that \[\min(d,\err_T)\leq \min(K,\ln K\sqrt{\epsilon^{-1}K\ln (T\ln K/\beta)}),\] and for $\delta > 0$ that \[\min(d,\err_T)\leq \min(K,\left(\epsilon^{-2}K\ln^2 K\ln(1/\delta)\ln^2(T/\beta)\right)^{1/3}). \qedhere \]
\end{proof}
\section{The Sparse Vector Technique}
\label{sec:sparsevector}

\newcommand{\T}{\mathrm{Thresh}}
\newcommand{\yes}{\textsc{yes}}
\newcommand{\no}{\textsc{no}}
The sparse vector technique is based on an algorithm in \cite{DBLP:conf/stoc/DworkNRRV09} and was described more fully in \cite{journals/fttcs/DworkR14}. The version described in Algorithm \ref{alg:sparsevector} is from \cite{journals/pvldb/LyuSL17} for $c=1$ (the main difference is that it allows different thresholds for every query).

\begin{algorithm}[tb]

\begin{algorithmic}[1]
\STATE{\bf Input:} Dataset $D$, threshold $\T$, and queries $q_1,q_2,\dots$ which have sensitivity at most $1$.
\STATE $\tau=\Lap(2/\epsilon)$\;

\FOR{$t=1,\dots,$ }
    \STATE $\mu_t=\Lap(4/\epsilon)$\;
    \IF{$q_t(D)+\mu_t>\T+\tau$}
        \STATE output $a_t=\yes$\;
        \STATE {\bf Abort}
        \ELSE
        \STATE output $a_t=\no$\;
        \ENDIF
\ENDFOR

\end{algorithmic}
\caption{AboveThreshold with output}
\label{alg:sparsevector}
\end{algorithm}

\begin{lemma}[\cite{journals/fttcs/DworkR14}]\label{lem:SVpriv}
    Algorithm \ref{alg:sparsevector} is $\epsilon$-differentially private.
\end{lemma}

\begin{lemma}[\cite{journals/fttcs/DworkR14}]\label{lem:SVacc}
    Algorithm \ref{alg:sparsevector} fulfills the following accuracy guarantees for $\alpha=\frac{8(\ln k + \ln(2/\beta))}{\epsilon}$:
    For any sequence $q_1,\dots,q_k$ of queries it holds with probability at least $1-\beta$,
    \begin{enumerate}
        \item for any $t$ such that $a_t=\yes$ we have
        \begin{align*}
            q_t(D)\geq \T-\alpha,
        \end{align*}
        \item for all $t$ such that $a_t=\no$ we have
        \begin{align*}
            q_t(D)\leq \T+\alpha.
        \end{align*}
    \end{enumerate}
\end{lemma}

\end{document}